\documentclass[twocolumn,pra,showpacs,superscriptaddress]{revtex4-1}
\usepackage{amssymb}
\usepackage{mathrsfs}
\usepackage[pdftex]{graphicx}
\usepackage[caption=false]{subfig}
\usepackage{mathtools}
\usepackage[pdftex,colorlinks=true]{hyperref}
\DeclareGraphicsExtensions{.pdf,.jpg,.png}
\usepackage[toc,page,title,titletoc,header]{appendix}
\usepackage{tikz}
\usetikzlibrary{decorations.pathmorphing}

\begin{document}

\title{Tunable coupling between a superconducting resonator and an
  artificial atom}

\author{Qi-Kai He}

\affiliation{Institute of Physics, Beijing National Laboratory for
  Condensed Matter Physics, Chinese Academy of Sciences, Beijing
  100190, China}

\affiliation{School of Physical Sciences, University of Chinese
  Academy of Sciences, Beijing 100049, China}

\author{D.~L. Zhou}

\email{zhoudl72@iphy.ac.cn}

\affiliation{Institute of Physics, Beijing National Laboratory for
  Condensed Matter Physics, Chinese Academy of Sciences, Beijing
  100190, China}

\affiliation{School of Physical Sciences, University of Chinese
  Academy of Sciences, Beijing 100049, China}

\date{\today}

\begin{abstract}
  Coherent manipulation of a quantum system is one of the main
  themes in current physics researches. In this work, we design a
  circuit QED system with a tunable coupling between an artificial
  atom and a superconducting resonator while keeping the cavity
  frequency and the atomic frequency invariant. By controlling the
  time dependence of the external magnetic flux, we show that it is
  possible to tune the interaction from the extremely weak coupling
  regime to the ultrastrong coupling one. Using the quantum
  perturbation theory, we obtain the coupling strength as a function
  of the external magnetic flux. In order to show its reliability in
  the fields of quantum simulation and quantum computing, we study
  its sensitivity to noises.
\end{abstract}


\maketitle

\section{Introduction}

The light-matter interaction in circuit quantum electrodynamics (QED)
finds lots of applications in many quantum information processes, such
as the simulation of resonance fluorescence~\cite{PhysRevX.6.031004},
an experimental proposal for boson sampling~\cite{BosonSampling2016}
and the single-photon scattering on an
atom~\cite{PhysRevLett.101.100501, He_singlephoton,
forn2017ultrastrong}. All these achievements show that circuit QED is
an excellent platform for studying the physics induced by light-matter
interaction~\cite{forndiaz2010observation, forn2017ultrastrong,
GU20171, yoshihara2016superconducting}.

In previous studies, the photon-photon
interaction~\cite{peropadre2013tunable, baust2015tunable,
  wulschner2016tunable} and the atom-atom
interaction~\cite{PhysRevB.74.184501} have been investigated with
elaborate superconducting Josephson junction circuits, and the
photon-atom interaction has been studied in all coupling regimes.
Nevertheless, we note that it is rarely mentioned how to control the
light-matter interaction during an adiabatic
process~\cite{tong2007sufficiency}, which is inevitable and valuable
in such experiments. To this end, it becomes an urgent need to design
a superconducting circuit for implementing the light-atom interaction
with a continuously tunable coupling strength and with a fixed qubit
frequency.

Designing such a circuit is equivalent to devising an unusual qubit.
The widely used superconducting qubits includes phase
qubit~\cite{PhysRevLett.95.210503}, transmon~\cite{koch2007charge},
and Xmon~\cite{barends2013coherent}, which are different assemblies of
Josephson junctions and other circuit components like capacitance,
inductance. Because of different structures of these qubits, they can
be manipulated by different external signals and sensitive to diverse
noise sources. In the circuit QED, the single mode cavity can be
realized by a LC oscillator or a microwave transmission line. To
construct a quantum network, we need to couple qubits and photons,
which can be implemented by the capacitance
connection~\cite{PhysRevB.86.140508}, the mutual
inductance~\cite{PhysRevB.78.104508, 1367-2630-7-1-230} and the direct
connection~\cite{PhysRevLett.105.023601}. In particular, it is worthy
to point out that the direct connection between a qubit and a
microwave transmission line has led the coupling strength into the
ultrastrong regime~\cite{PhysRevLett.105.023601, forn2017ultrastrong}.

Up to now, the qubit-resonator coupling strength can be tuned from
zero to a finite value~\cite{PhysRevLett.106.030502,
  srinivasan2011tunable, bruschi2013time, mezzacapo2014many}. This
coupling strength is controlled by the direct capacitance between the
upper island and the lower island, which can't be tuned in time. Other
than directly changing the capacitance, the external signals like
magnetic flux can also be used to modify the photon-atom interaction.
In general, the qubit frequency will be changed with the variation of
the coupling strength~\cite{forn2017ultrastrong}. If we ignore the
tiny variation of the qubit frequency, the dc-SQUID can be used as a
tunable coupler to modulate the coupling
strength~\cite{PhysRevLett.119.150502}. However, the coupling strength
controlled by this circuit design cannot be tuned in all coupling
regimes.

Inspired by Refs.~\cite{PhysRevB.86.140508} and
\cite{PhysRevLett.105.023601}, we devise an artificial atom based on
the superconducting loop containing two dc-SQUIDs and make the
photon-atom coupling flux-controlled. To see clearly the underlying
mechanism, we theoretically calculate the coupling strength between
two coupled states. As an application, we investigate the adiabatic
dynamics of the qubit in superconducting resonators in all coupling
regimes. Furthermore, we discuss the influence of the resistance of
Josephson junctions on our circuit and show its extensive
applicability in quantum simulations. For example, our system allows
the adiabatic switching of the coupling strength from the strong to
the ultra strong coupling regime which makes it a reliable tool for
implementing a quantum memory~\cite{kyaw2015scalable}. Moreover, if
fast-tuning is possible, our system can also be used for simulating
the relativistic effects~\cite{PhysRevB.92.064501,
garciaalvarez2017entanglement}.

The rest of this paper is organized as follows. In Sec.~\ref{sec:2},
after introducing our design, we analytically study the Hamiltonian of
the artificial atom. With the help of quantum perturbation theory, we
obtain the explicit formula of the coupling strength and the energy
difference between two lowest eigenstates of the atom, which are
detailed derived in Appendix~\ref{app:A}. Based on the formula, we
propose the tunable coupling scheme, which makes our system a tunable
coupling Rabi model. To test the experimental feasibility of our
circuit, we consider the dissipation of the artificial atom in
Sec.~\ref{sec:3}. With all these analytical results, we estimate the
proper value of device parameters of our qubit in a circuit QED
experiment in Sec.~\ref{sec:4}. In Sec.~\ref{sec:5}, we draw the
conclusions.

\section{Tunable coupling\label{sec:2}}

In this section, we study the mechanism for the coupling between an
artificial atom and the superconducting resonator. The artificial atom
is based on superconducting quantum interference devices (SQUIDs),
which are widely used in circuit QED\@. To manipulate the energy
splitting of the system as well as the coupling strength with
outsides, we control three time-dependent magnetic fluxes. Since the
coupling strength need to be tuned into the ultrastrong regime, we
directly connect the atom to the transmission line. Based on the two
previous ideas, we theoretically design the artificial atom
schematically shown in Fig.~\ref{fig:Loop_SQUID}. We will give its
Hamiltonian and derive the expression for the coupling strength.

\subsection{System Hamiltonian\label{sec:2A}}

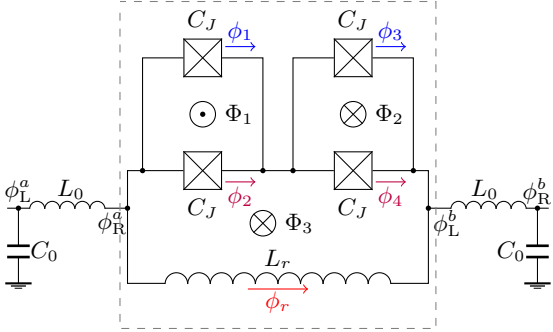
\begin{figure}[!htbp]
  \centering
  \begin{tikzpicture}[node-s1/.style = {rectangle, draw, inner
      sep=0.25cm, fill = white}, node-s2/.style = {circle, draw, inner
      sep=0.12cm, fill = white}]
    \node[node-s1, label = above: $C_{J}$] (j1) at (-1, 1.5) {};
    \node[node-s1, label = below: $C_{J}$] (j2) at (-1, 0) {};
    \node[node-s1, label = above: $C_{J}$] (j3) at (1, 1.5) {};
    \node[node-s1, label = below: $C_{J}$] (j4) at (1, 0) {};
    \node[node-s2, label = right: $\Phi_{1}$] (c1) at (-1, 0.75) {};
    \node[node-s2, label = right: $\Phi_{2}$] (c2) at (1, 0.75) {};
    \draw[-] (j1.north west) -- (j1.south east);
    \draw[-] (j1.north east) -- (j1.south west);
    \draw[-] (j2.north west) -- (j2.south east);
    \draw[-] (j2.north east) -- (j2.south west);
    \draw[-] (j3.north west) -- (j3.south east);
    \draw[-] (j3.north east) -- (j3.south west);
    \draw[-] (j4.north west) -- (j4.south east);
    \draw[-] (j4.north east) -- (j4.south west);
    \draw[-] (c2.north west) -- (c2.south east);
    \draw[-] (c2.north east) -- (c2.south west);
    \fill[black] (-1, 0.75) circle(1pt);
    \draw[-] (-1.5, -1.5) -- (-2, -1.5) |- (j2) -- (j4) -| (2, -1.5)
    -- (1.5, -1.5) (-1.8, 0) |- (j1) -| (-0.2, 0) (0.2, 0) |- (j3)
    -| (1.8, 0);
    \fill[black] (-1.8, 0) circle(1pt);
    \fill[black] (-0.2, 0) circle(1pt);
    \fill[black] (0.2, 0) circle(1pt);
    \fill[black] (1.8, 0) circle(1pt);
    \fill[black] (-2, -0.5) circle(1pt);
    \fill[black] (2, -0.5) circle(1pt);
    \draw (-1.2,-1.5-0.00875) arc (0 : 180 : 0.15);
    \draw (-0.9,-1.5-0.00875) arc (0 : 180 : 0.15);
    \draw (-0.6,-1.5-0.00875) arc (0 : 180 : 0.15);
    \draw (-0.3,-1.5-0.00875) arc (0 : 180 : 0.15);
    \draw (0,-1.5-0.00875) arc (0 : 180 : 0.15);
    \draw (0.3,-1.5-0.00875) arc (0 : 180 : 0.15);
    \draw (0.6,-1.5-0.00875) arc (0 : 180 : 0.15);
    \draw (0.9,-1.5-0.00875) arc (0 : 180 : 0.15);
    \draw (1.2,-1.5-0.00875) arc (0 : 180 : 0.15);
    \draw (1.5,-1.5-0.00875) arc (0 : 180 : 0.15);
    \node at (0, -1.175) {$L_{r}$};
    \draw[->, blue] (-0.7, 1.65) -- (-0.3, 1.65);
    \draw[->, purple] (-0.7, -0.15) -- (-0.3, -0.15);
    \draw[<-, blue] (1.7, 1.65) -- (1.3, 1.65);
    \draw[<-, purple] (1.7, -0.15) -- (1.3, -0.15);
    \node at (-0.5, 1.85) {$\textcolor{blue}{\phi_{1}}$};
    \node at (-0.5, -0.35) {$\textcolor{purple}{\phi_{2}}$};
    \node at (1.5, 1.85) {$\textcolor{blue}{\phi_{3}}$};
    \node at (1.5, -0.35) {$\textcolor{purple}{\phi_{4}}$};
    \draw[->, red] (-0.4, -1.575) -- (0.4, -1.575);
    \node at (0, -1.75) {$\textcolor{red}{\phi_{r}}$};
    \node[node-s2, label = right: $\Phi_{3}$] (cf) at (-0.2, -0.7) {};
    \draw[-] (cf.north west) -- (cf.south east);
    \draw[-] (cf.north east) -- (cf.south west);
    \draw[-] (-2.3,-0.5) -- (-2,-0.5) (2,-0.5) -- (2.3,-0.5) (-3.3,-0.5) -- (-3.6,-0.5) (3.3,-0.5) -- (3.6,-0.5) (-3.45,-0.5) -- (-3.45,-1) (-3.45,-1.15) -- (-3.45,-1.5) (3.45,-0.5) -- (3.45,-1) (3.45,-1.15) -- (3.45,-1.5);
    \draw[-, line width = 1pt] (-3.6,-1) -- (-3.3,-1) (-3.6,-1.15) -- (-3.3,-1.15) (3.6,-1) -- (3.3,-1) (3.6,-1.15) -- (3.3,-1.15);
    \draw[-, line width = 0.8pt] (-3.625,-1.5) -- (-3.275,-1.5) (3.275,-1.5) -- (3.625,-1.5);
    \draw[-, line width = 0.5pt] (-3.55,-1.535) -- (-3.35,-1.535) (3.55,-1.535) -- (3.35,-1.535);
    \draw[-, line width = 0.3pt] (-3.5,-1.565) -- (-3.4,-1.565) (3.4,-1.565) -- (3.5,-1.565);
    \draw (-3.1,-0.50875) arc (0 : 180 : 0.1);
    \draw (-2.9,-0.50875) arc (0 : 180 : 0.1);
    \draw (-2.7,-0.50875) arc (0 : 180 : 0.1);
    \draw (-2.5,-0.50875) arc (0 : 180 : 0.1);
    \draw (-2.3,-0.50875) arc (0 : 180 : 0.1);
    \draw (2.5,-0.50875) arc (0 : 180 : 0.1);
    \draw (2.7,-0.50875) arc (0 : 180 : 0.1);
    \draw (2.9,-0.50875) arc (0 : 180 : 0.1);
    \draw (3.1,-0.50875) arc (0 : 180 : 0.1);
    \draw (3.3,-0.50875) arc (0 : 180 : 0.1);
    \fill[black] (-3.45,-0.5) circle(1pt);
    \fill[black] (3.45,-0.5) circle(1pt);
    \node at (-2.75,-0.25) {$L_{0}$};
    \node at (2.75,-0.25) {$L_{0}$};
    \node at (-3.1,-1.075) {$C_{0}$};
    \node at (3.1,-1.075) {$C_{0}$};
    \node at (-2.2,-0.7) {$\phi^{a}_{\rm{R}}$};
    \node at (2.25,-0.7) {$\phi^{b}_{\rm{L}}$};
    \node at (-3.45,-0.275) {$\phi^{a}_{\rm{L}}$};
    \node at (3.45,-0.275) {$\phi^{b}_{\rm{R}}$};
    \draw[-, gray, dashed] (-2.1,2.1) |- (2.1,-2.1) |- (-2.1,2.25);
  \end{tikzpicture}
  \caption{(Color online). The schematic of the circuit layout of an
    artificial atom sharing an inductance with a coplanar transmission
    line. In the lumped element approximation, this transmission line
    resonator is constructed by two LC oscillators with the inductance
    $L_{0}$ and capacitance $C_{0}$. The artificial atom in the dashed
    gray rectangle is composed of a superconducting loop with two
    dc-SQUIDs. $C_{J}$ represents the capacitance of every Josephson
    junction. Three external magnetic fluxes are denoted as
    $\Phi_{1},\Phi_{2}$ and $\Phi_{3}$. $\phi^{a(b)}_{\rm{L}}$
    and $\phi^{a(b)}_{\rm{R}}$ are the reduced node flux for each
    side of the LC resonator.}
  \label{fig:Loop_SQUID}
\end{figure}

In Fig.~\ref{fig:Loop_SQUID}, an artificial atom is galvanically
attached to the center of the coplanar waveguide transmission line
resonator which is described as a two-mode LC resonator with its
inductance $L_{0}$ and capacitance $C_{0}$. As shown in the dashed
gray rectangle of Fig.~\ref{fig:Loop_SQUID}, this atom contains
four Josephson junctions that are designed with the same area and
combined into two dc-SQUIDs. Similar to the flux qubit, our atom
adopts a superconducting coil to connect dc-SQUIDs in series. In order
to eliminate the induced currents flowing to the connecting loop with
self-inductance $L_{r}$, the penetrating fluxes of the two dc-SQUIDs
have the same value ($\Phi_{1} = \Phi_{2}$) but are in opposite
directions.

For simplicity, we neglect the additional flux generated by the
circulating loop current in the dc-SQUIDs, which is equivalent to
restricting our discussion to the screening parameter
$\beta_{L} = \dfrac{ L_{S} I_{c}}{\varphi_{0}} \ll 1$. Here, $L_{S}$
is the loop inductance of the dc-SQUID, $I_{c}$ is the critical
current of the Josephson junctions, and $\varphi_{0} = \dfrac{\hbar}{2
e}$ is the reduced flux quantum.

As shown in Fig.~\ref{fig:Loop_SQUID}, we divide the system into two
parts: the circuits inside and those outside of the dashed rectangle.
The Lagrangian of our system is
\begin{equation}
  \hat{L} = \hat{L}' + \hat{L}''
  \label{eq:2.A.1}
\end{equation}
with
\begin{subequations}
  \begin{align}
  & \hat{L}' = \sum^{4}_{i = 1} \left[\dfrac{1}{2} C_{J} \varphi^{2}_{0}
  \dot{\phi}^{2}_{i} - E_{J} (1 - \cos \phi_{i})\right] -
  \varphi_{0}^{2} \dfrac{(\phi^{b}_{\rm{L}} - \phi^{a}_{\rm{R}})^{2}}{2 L_{r}},
  \label{eq:2.A.2a} \\
  & \hat{L}'' = \dfrac{C_{0}}{2} \varphi^{2}_{0}
  [(\dot{\phi}^{a}_{\rm{L}})^{2} + (\dot{\phi}^{b}_{\rm{R}})^{2}] -
  \varphi^{2}_{0} \dfrac{(\phi^{a}_{\rm{L}} - \phi^{a}_{\rm{R}})^{2} +
  (\phi^{b}_{\rm{L}} - \phi^{b}_{\rm{R}})^{2}}{2 L_{0}}, \label{eq:2.A.2b}
  \end{align}
  \label{eq:2.A.2}
\end{subequations}
where $L^{\prime}$ and $L^{\prime\prime}$ are the Lagrangian of the
inside part and that of the outside part respectively, $\phi_{i}~(i =
1,2,3,4)$ is the phase difference across the $i$-th junction, and
$\phi^{a(b)}_{\rm{L(R)}} = \dfrac{1}{\varphi_{0}} \int^{t}_{-\infty}
V^{a(b)}_{\rm{L(R)}} (\tau) d \tau$ is the reduced node
flux~\cite{DevoretQuanFluc} which corresponds to the electric
potential $V^{a(b)}_{\rm{L(R)}}$ for each side of the LC resonator.

Following Ref.~\cite{PhysRevLett.105.023601, PhysRevLett.111.243602},
we introduce $\varphi_{\pm} = \phi^{b}_{\rm{R}} \pm
\phi^{a}_{\rm{L}}$, $\Phi_{+} = \phi^{b}_{\rm{L}} + \phi^{a}_{\rm{R}}$
and consider $\phi_{r} = \phi^{b}_{\rm{L}} - \phi^{a}_{\rm{R}}$, then
we rewrite the Lagrangian~\eqref{eq:2.A.2a} and \eqref{eq:2.A.2b} into
\begin{subequations}
  \begin{align}
  & \hat{L}' = \sum^{4}_{i = 1} \left[\dfrac{1}{2} C_{J}
  \varphi^{2}_{0} \dot{\phi}^{2}_{i} - E_{J} (1 - \cos
  \phi_{i})\right] - \varphi_{0}^{2} \dfrac{\phi_{r}^{2}}{2 L_{r}},
  \label{eq:2.A.3a} \\
  & \hat{L}'' = \dfrac{C_{0}}{4} \varphi^{2}_{0}
  (\dot{\varphi}^{2}_{+} + \dot{\varphi}^{2}_{-}) - \varphi^{2}_{0}
  \dfrac{(\varphi_{+} - \Phi_{+})^{2} + (\varphi_{-} -
  \phi_{r})^{2}}{4 L_{0}}.
  \label{eq:2.A.3b}
  \end{align}
  \label{eq:2.A.3}
\end{subequations}

Note that the conditions of fluxoid quantization along the independent
loops in the circuit are given by
\begin{subequations}
  \begin{align}
    & \phi_{2} - \phi_{1} = f, \label{eq:2.A.5a} \\
    & \phi_{3} - \phi_{4} = f, \label{eq:2.A.5b} \\
    & \phi_{2} + \phi_{4} - \phi_{r} = f', \label{eq:2.A.5c}
  \end{align}
  \label{eq:2.A.4}
\end{subequations}
where
\begin{subequations}
  \begin{align}
    & f =  \dfrac{\Phi_{1}}{\varphi_{0}} =
    \dfrac{\Phi_{2}}{\varphi_{0}}, \label{eq:2.A.5a} \\
    & f' =  \dfrac{\Phi_{3}}{\varphi_{0}}. \label{eq:2.A.5b}
  \end{align}
  \label{eq:2.A.5}
\end{subequations}

With the Lagrangian~\eqref{eq:2.A.3} and the
relation~\eqref{eq:2.A.4}, we write the Hamiltonian of our system
\begin{equation}
  \begin{split}
    & \hat{H}' = \dfrac{\Omega^{2}_{J} p^{2}_{+}}{8 \hbar \omega_{r}}
    + \dfrac{\Omega^{2}_{J} p^{2}_{-}}{8 \hbar \omega_{r}} + 2 \hbar
    \omega_{r} \phi^{2}_{+} + 2 E_{J} \left\{2 - \cos \dfrac{f}{2}
    \right.\\
     & \left.\times \left[\cos (\phi_{+} + \phi_{-} - \dfrac{f -
     f'}{2}) + \cos (\phi_{+} - \phi_{-} + \dfrac{f +
     f'}{2})\right]\right\},
  \end{split}
  \label{eq:2.A.6}
\end{equation}
and
\begin{equation}
  \begin{split}
    \hat{H}'' = & \dfrac{\omega^{2}_{c} p'^{2}_{+}}{\hbar \omega_{0}}
    + \dfrac{\hbar \omega_{0} \varphi^{2}_{+}}{4} + \dfrac{\hbar
    \omega_{0} \Phi^{2}_{+}}{4} - \dfrac{\hbar \omega_{0}}{2}
    \varphi_{+} \Phi_{+} \\
    & + \dfrac{\omega^{2}_{c} p'^{2}_{-}}{\hbar \omega_{0}} +
    \dfrac{\hbar \omega_{0} \varphi^{2}_{-}}{4} + \hbar \omega_{0}
    \phi^{2}_{+} - \hbar \omega_{0} \varphi_{-} \phi_{+},
  \end{split}
  \label{eq:2.A.7}
\end{equation}
where $\Omega_{J} = \dfrac{1}{\sqrt{L_{r} C_{J}}}$, $\hbar \omega_{r}
= \dfrac{\varphi^{2}_{0}}{L_{r}}$, $\omega_{c} = \dfrac{1}{\sqrt{L_{0}
C_{0}}}$, $\hbar \omega_{0} = \dfrac{\varphi^{2}_{0}}{L_{0}}$, and
$p_{\pm} = 4 C_{J} \varphi^{2}_{0} \dot{\phi}_{\pm}$ is the conjugate
momentum with corresponding phase difference $\phi_{\pm} =
\dfrac{(\phi_{2} - f'/2) \pm (\phi_{4} - f'/2)}{2}$, $p'_{\pm} =
\dfrac{1}{2} C_{0} \varphi^{2}_{0} \dot{\varphi}_{\pm}$ is the
canonical momentum which corresponds to $\varphi_{\pm}$.

From Eqs.~\eqref{eq:2.A.6} and \eqref{eq:2.A.7}, we can indicate that
$\hat{H}'$ represents the Hamiltonian of the artificial atom without
outside connections, $\hat{H}''$ describes the two-mode LC resonator
with intrinsic frequency $\omega_{c}$. Since $\phi_{+}$ forms the part
of the qubit while $\Phi_{+}$ does not, we conclude that the qubit
only couples with one mode of the resonator and $- \hbar \omega_{0}
\varphi_{-} \phi_{+}$ represents the qubit-resonator interaction. In
this sense, we can treat the circuit shown in
Fig.~\ref{fig:Loop_SQUID} as a coupling system with one atom and a
single-mode cavity. In this simplified system, the atomic Hamiltonian
should consider not only the inside part of the dashed rectangle but
also the renormalization $\hbar \omega_{0} \phi^{2}_{+}$ from
outsides. Then the Hamiltonian of our system is
\begin{equation}
  \hat{H} = \hat{H}_{\rm{atom}} + \hat{H}_{\rm{cav}} +
  \hat{H}_{\rm{int}},
  \label{eq:2.A.8}
\end{equation}
where
\begin{subequations}
  \begin{align}
    & \hat{H}_{\rm{atom}} = \hat{H}' + \hbar \omega_{0} \phi^{2}_{+},
    \label{eq:2.A.9a} \\
    & \hat{H}_{\rm{cav}} = \dfrac{\omega^{2}_{c} p'^{2}_{-}}{\hbar
    \omega_{0}} + \dfrac{\hbar \omega_{0} \varphi^{2}_{-}}{4},
    \label{eq:2.A.9b} \\
    & \hat{H}_{\rm{int}} = - \hbar \omega_{0} \varphi_{-} \phi_{+}.
    \label{eq:2.A.9c}
  \end{align}
  \label{eq:2.A.9}
\end{subequations}

In the second quantization representation, the cavity
Hamiltonian~\eqref{eq:2.A.9b} is rewritten as
\begin{equation}
  \hat{H}_{\rm{cav}} = (\hat{a}^{\dagger} \hat{a} + \dfrac{1}{2})
  \hbar \omega_{c},
  \label{eq:2.A.10}
\end{equation}
where $\hat{a} = \sqrt{\dfrac{\omega_{0}}{4 \omega_{c}}}
\left(\hat{\varphi}_{-} + \dfrac{2 i \omega_{c}}{\hbar \omega_{0}}
\hat{p}'_{-}\right)$ is the photon annihilation operator for the
cavity. Then the interaction Hamiltonian~\eqref{eq:2.A.9c} can be
changed to
\begin{equation}
  \hat{H}_{\rm{int}} = - \hbar \sqrt{\omega_{0} \omega_{c}} \phi_{+}
  (\hat{a}^{\dagger} + \hat{a}),
  \label{eq:2.A.11}
\end{equation}
which implies that the tunable coupling to the cavity is mainly
determined by $\phi_{+}$.

\subsection{Theoretical analysis of the coupling
strength\label{sec:2B}}

To give a theoretical analysis of the coupling strength, we resort to
the quantum perturbation theory~\cite{Sakurai2011Modern}. Now we
consider the case of $f=\pi - \Delta$ ($\Delta \ll \pi$). For
simplicity, we define the charging energy $E_{c} = \dfrac{e^{2}}{2
C_{J}}$ and $\hbar \omega'_{r} = \dfrac{\varphi^{2}_{0}}{L'_{r}}$ with
$L'_{r} = \dfrac{2 L_{0} L_{r}}{2 L_{0} + L_{r}}$. Then we divide the
Hamiltonian~\eqref{eq:2.A.9a} into two parts, $\hat{H}_{\rm{atom}} =
\hat{H}_{0} + \hat{V}$, where the unperturbed Hamiltonian
\begin{equation}
  \hat{H}_{0} = \dfrac{E_{c}}{\hbar^{2}} \hat{p}^{2}_{+} + 2 \hbar
  \omega'_{r} \phi^{2}_{+} + \dfrac{E_{c}}{\hbar^{2}} \hat{p}^{2}_{-}
  + 4 E_{J},
  \label{eq:2.B.1}
\end{equation}
and a perturbation term
\begin{equation}
  \hat{V} = - 2 \Delta E_{J} \cos (\phi_{+} + \dfrac{f'}{2}) \sin
  (\phi_{-} + \dfrac{\Delta}{2})
  \label{eq:2.B.2}
\end{equation}
with $\Delta$ being a small parameter.

In the unperturbed Hamiltonian~\eqref{eq:2.B.1}, the conjugate
momentum $\hat{p}_{+}$ and its corresponding coordinate $\phi_{+}$
describe a harmonic oscillator, and the other canonical momentum
$\hat{p}_{-} = \hat{n}_{-} \hbar$ with $\hat{n}_{-}$ being the
relative cooper pair number operator between the two dc-SQUIDs. In
other words, the unperturbed Hamiltonian given by Eq.~\eqref{eq:2.B.1}
can be rewritten as
\begin{equation}
  \hat{H}_{0} = E_{b} (\hat{b}^{\dagger} \hat{b} + \dfrac{1}{2})  + E_{c} \hat{n}^{2}_{-} + 4 E_{J},
  \label{eq:2.B.3}
\end{equation}
where $E_{b}=\sqrt{8\hbar \omega'_{r} E_{c}}$, $\hat{b} = \dfrac{1}{2
\lambda} \left(\hat{\phi}_{+} + \dfrac{2 i \lambda^{2}}{\hbar}
\hat{p}_{+}\right)$ is the annihilation operator with $\lambda =
\sqrt{\dfrac{E_{c}}{E_{b}}}$. We solve the eigen problem of
$\hat{H}_{0}$:
\begin{equation}
  \hat{H}_{0} |n; n_{-}\rangle = E^{(0)}_{n;n_{-}} |n; n_{-}\rangle,
  \label{eq:2.B.4}
\end{equation}
where $\hat{b}^{\dagger} \hat{b} | n \rangle = n | n \rangle$,
$\hat{n}_{-} | n_{-} \rangle = n_{-} | n_{-} \rangle$, and
\begin{equation}
  E^{(0)}_{n; n_{-}} = (n + \dfrac{1}{2}) E_{b} + E_{c}
  n^{2}_{-} + 4 E_{J}.
  \label{eq:2.B.5}
\end{equation}
It is easy to see that all excited states with $n_{-}\neq 0$ of
$\hat{H}_{0}$ are two-fold degenerate. In our paper, we focus on the
region of $E_{b} \gg E_{c}$, where the lower energy eigen states
satisfies $n=0$, e.g., $|0; 0\rangle$ is the ground state, and $|0;
\pm 1\rangle$ are the lowest degenerate excited states.

To determine the eigenstates of $\hat{H}_{\text{atom}}$, we first
perform the symmetry analysis of the Hamiltonian. In fact, the
Hamiltonian is invariant under the transformation $\phi_{-} \to f -
\phi_{-}$, which corresponds to the parity operator
\begin{equation}
  \begin{split}
    \hat{P} = & \int d \phi_{-} | f - \phi_{-} \rangle \langle
    \phi_{-} | \\
    = & \sum_{n_{-}} \sum_{n'_{-}} \int d \phi_{-} | n_{-} \rangle
    \langle n_{-} | \pi - \phi_{-} - \Delta \rangle \langle \phi_{-} |
    n'_{-} \rangle \langle n'_{-} | \\
    = & \dfrac{1}{2 \pi} \sum_{n_{-}} \sum_{n'_{-}} \int d \phi_{-} |
    n_{-} \rangle e^{-i n_{-} (\pi - \phi_{-} - \Delta)} e^{i n'_{-}
    \phi_{-}} \langle n'_{-} | \\
    = & \sum_{n_{-}} | n_{-} \rangle e^{-i n_{-} (\pi - \Delta)}
    \langle - n_{-} |,
  \end{split}
  \label{eq:2.B.6}
\end{equation}
where $|\phi_{-}\rangle$ is the eigenstate of $\hat{\phi}_{-}$ with
the eigenvalue $\phi_{-}$. Actually, it is easy to check that
$[\hat{P}, \hat{H}_{0}]=[\hat{P}, \hat{V}]=[\hat{P},
\hat{H}_{\rm{atom}}] = 0$. Therefore we can always choose the
eigenstate of $\hat{H}_{\text{atom}}$ with definite parity, whose
zero-order eigenstate has the same parity. The eigen problem of
$\hat{P}$ is given by for any positive integer $n_{-}$
\begin{equation}
  \hat{P} |\pm_{n_{-}}\rangle  = \pm |\pm_{n_{-}}\rangle,
  \label{eq:2.B.7}
\end{equation}
where
\begin{align}
  |\pm_{n_{-}}\rangle & = e^{i n_{-} \Delta/2}\frac{|n_{-}\rangle \pm
  \hat{P} |n_{-}\rangle}{\sqrt{2}} \nonumber\\
  & = \frac{e^{i n_{-} \Delta/2} |n_{-}\rangle \pm (-1)^{n_{-}} e^{-i
  n_{-} \Delta/2} |- n_{-}\rangle}{\sqrt{2}}.
  \label{eq:2.B.8}
\end{align}
It is worthy to note that the state $|n_{-}=0\rangle$ is with even
parity: $\hat{P} | 0 \rangle = | 0 \rangle$.

Using the parity operator $\hat{P}$, we write the zero-order
eigenstates of two lowest excited states as
\begin{subequations}
  \begin{align}
    & | \psi^{(0)}_{+} \rangle = |0; +_{1}\rangle, \label{eq:2.B.9a}
    \\
    & | \psi^{(0)}_{-} \rangle = |0; -_{1}\rangle \label{eq:2.B.9b}
  \end{align}
  \label{eq:2.B.9}
\end{subequations}
which obeys $\hat{P} | \psi^{(0)}_{+} \rangle = | \psi^{(0)}_{+}
\rangle$ and $\hat{P} | \psi^{(0)}_{-} \rangle = - | \psi^{(0)}_{-}
\rangle$, that is, $| \psi^{(0)}_{+} \rangle$ and the zero-order
ground state $| \psi^{(0)}_{0} \rangle = |0; 0 \rangle$ are the states
with even parity and $| \psi^{(0)}_{-} \rangle$ is that with odd
parity. Then the three lowest eigenstates of $\hat{H}_{\text{atom}}$,
$| \psi_{0} \rangle$, $| \psi_{+} \rangle$, and $| \psi_{-} \rangle$
have the same parity as their zero-order eigenstates $| \psi_{0}^{(0)}
\rangle$, $| \psi_{+}^{(0)} \rangle$, and $| \psi_{-}^{(0)} \rangle$
respectively.

Since $[\hat{\phi}_{+}, \hat{P}]  = 0$, we have
\begin{subequations}
  \begin{align}
    & \langle \psi_{0} | \hat{\phi}_{+} | \psi_{-} \rangle = \langle
    \psi_{0} | \hat{P} \hat{\phi}_{+} \hat{P} | \psi_{-} \rangle = -
    \langle \psi_{0} | \hat{\phi}_{+} | \psi_{-} \rangle,
    \label{eq:2.B.10a} \\
    & \langle \psi_{+} | \hat{\phi}_{+} | \psi_{-} \rangle = \langle
    \psi_{+} | \hat{P} \hat{\phi}_{+} \hat{P} | \psi_{-} \rangle = -
    \langle \psi_{+} | \hat{\phi}_{+} | \psi_{-} \rangle,
    \label{eq:2.B.10b}
  \end{align}
  \label{eq:2.B.10}
\end{subequations}
which implies that $\langle \psi_{0} | \hat{\phi}_{+} | \psi_{-}
\rangle = \langle \psi_{+} | \hat{\phi}_{+} | \psi_{-} \rangle
= 0$. Hence we can restrict the Hilbert space of the artificial atom
into the subspace with the bases
$\{ | g \rangle = | \psi_{0} \rangle, | e \rangle = | \psi_{+} \rangle\}$. In this subspace, the
Hamiltonians
\begin{subequations}
  \begin{align}
    \hat{H}_{\text{atom}} & = E_{g} | g \rangle \langle g | + E_{e} |
    e \rangle \langle e |, \label{eq:2.B.11a}\\
    \hat{H}_{\text{int}} & = - \hbar \sqrt{\omega_{0} \omega_{c}}
    \left(\langle e|\hat{\phi}_{+}|g\rangle |e\rangle \langle g| +
    \langle g|\hat{\phi}_{+}|e\rangle |g\rangle \langle e| \right.
    \nonumber \\
    & \quad \left. + \langle g|\hat{\phi}_{+}|g \rangle |g\rangle
    \langle g| + \langle e|\hat{\phi}_{+}|e\rangle |e\rangle \langle
    e|\right) (\hat{a}^{\dagger} + \hat{a}), \label{eq:2.B.11b}
  \end{align}
  \label{eq:2.B.11}
\end{subequations}
where
\begin{equation}
  E_{g} \approx \frac{1}{2} E_{b} + 4 E_{J} - 2 \Delta^{2} E_{J}^{2}
  e^{-\lambda^{2}} \left(\dfrac{\cos^{2} \dfrac{f^{\prime}}{2}}{E_{c}}
  + \dfrac{\lambda^{2} \sin^{2} \dfrac{f^{\prime}}{2}}{E_{b}}\right),
  \label{eq:2.B.12}
\end{equation}
\begin{equation}
  \begin{split}
    E_{e} \approx & \frac{1}{2} E_{b} + E_{c} + 4 E_{J} + \dfrac{5}{3}
    \Delta^{2} \dfrac{E_{J}^{2}}{E_{c}} e^{-\lambda^{2}} \cos^{2}
    \dfrac{f^{\prime}}{2} \\
    & - 3 \Delta^{2} \dfrac{E^{2}_{J}}{E_{b}} \lambda^{2} \sin^{2}
    \dfrac{f^{\prime}}{2},
  \end{split}
  \label{eq:2.B.13}
\end{equation}
and
\begin{subequations}
  \begin{align}
    \langle e | \hat{\phi}_{+} | g \rangle & = (\langle g |
    \hat{\phi}_{+} | e \rangle)^{\ast} \approx i 2 \sqrt{2} \Delta
    \lambda^{2} \dfrac{E_{J}}{E_{b}} e^{- \lambda^{2}/2} \sin
    \dfrac{f'}{2}, \label{eq:2.B.14a} \\
    \langle g | \hat{\phi}_{+} | g \rangle & \approx - \dfrac{4
    \Delta^{2} E^{2}_{J} \lambda^{2} e^{- \lambda^{2}}}{E_{b} E_{c}}
    \sin f', \label{eq:2.B.14b} \\
    \langle e | \hat{\phi}_{+} | e \rangle & \approx \dfrac{10
    \Delta^{2} E^{2}_{J} \lambda^{2} e^{- \lambda^{2}}}{3 E_{b} E_{c}}
    \sin f'. \label{eq:2.B.14c}
  \end{align}
  \label{eq:2.B.14}
\end{subequations}

Finally we rewrite the Hamiltonian~\eqref{eq:2.A.8} as
\begin{equation}
  \hat{H} = (\hat{a}^{\dagger} \hat{a} + \dfrac{1}{2}) \hbar
  \omega_{c} + \dfrac{\delta E}{2} \hat{\sigma}_{z} + \hbar (g
  \hat{\sigma}_{y} + g_{0} \hat{\sigma}_{0} + g_{z} \hat{\sigma}_{z})
  (\hat{a}^{\dagger} + \hat{a}),
  \label{eq:2.B.15}
\end{equation}
where the Pauli operators $\hat{\sigma}_{z} = | e \rangle \langle e | - |g\rangle\langle g|$,
$\hat{\sigma}_{y}=-i|e\rangle\langle g|+i|g\rangle\langle e|$, $\hat{\sigma}_{0}=|e\rangle\langle e|+|g\rangle\langle g|$,
\begin{equation}
  \begin{split}
    \delta E = & E_{e} - E_{g} \\
    \approx & E_{c} + \Delta^{2} E_{J}^{2} e^{-\lambda^{2}}
    \left(\frac{11}{3 E_{c}} \cos^{2} \frac{f^{\prime}}{2} -
    \dfrac{\lambda^{2}}{E_{b}} \sin^{2} \dfrac{f'}{2}\right)
  \end{split}
  \label{eq:2.B.16}
\end{equation}
is the energy level splitting of the atom,
\begin{equation}
  g \approx \sqrt{8 \omega_{0} \omega_{c}} \Delta \lambda^{2}
  \dfrac{E_{J}}{E_{b}} e^{- \lambda^{2}/2} \sin \dfrac{f'}{2}
  \label{eq:2.B.17}
\end{equation}
and
\begin{subequations}
  \begin{align}
    g_{0} & \approx g \dfrac{\sqrt{2} \Delta E_{J}}{6 E_{c}} e^{-
    \lambda^{2}/2} \cos \dfrac{f'}{2}, \label{eq:2.B.18a} \\
    g_{z} & \approx - g \dfrac{11 \sqrt{2} \Delta E_{J}}{6 E_{c}}
    e^{- \lambda^{2}/2} \cos \dfrac{f'}{2} \label{eq:2.B.18b}
  \end{align}
  \label{eq:2.B.18}
\end{subequations}
are the coupling strength which corresponds to $\hat{\sigma}_{y}$,
$\hat{\sigma}_{0}$ and $\hat{\sigma}_{z}$ respectively (see detailed
derivations in Appendix~\ref{app:A}).
\begin{figure*}[!htbp]
  \centering
  \subfloat[$f'(f)$.]{
    \includegraphics[width=0.4\textwidth]{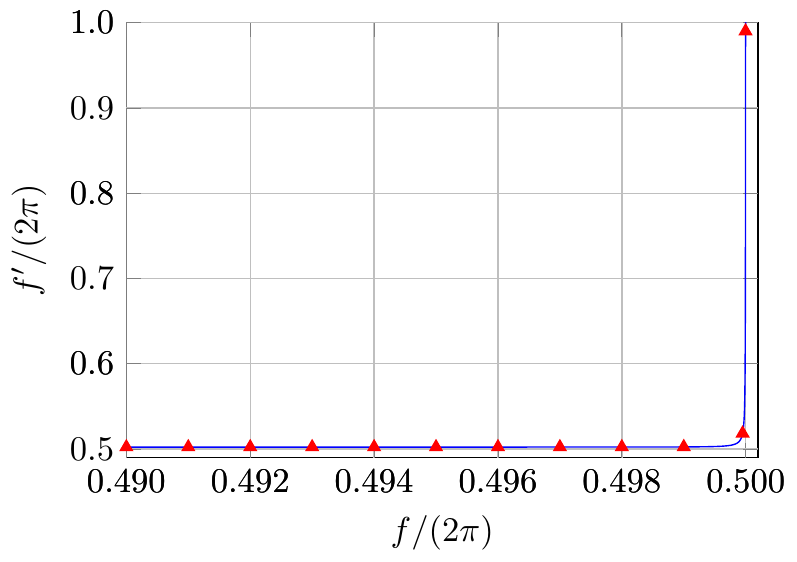}
    \label{fig:ff}
  }
  \subfloat[$E(f)$.]{
    \includegraphics[width=0.4\textwidth]{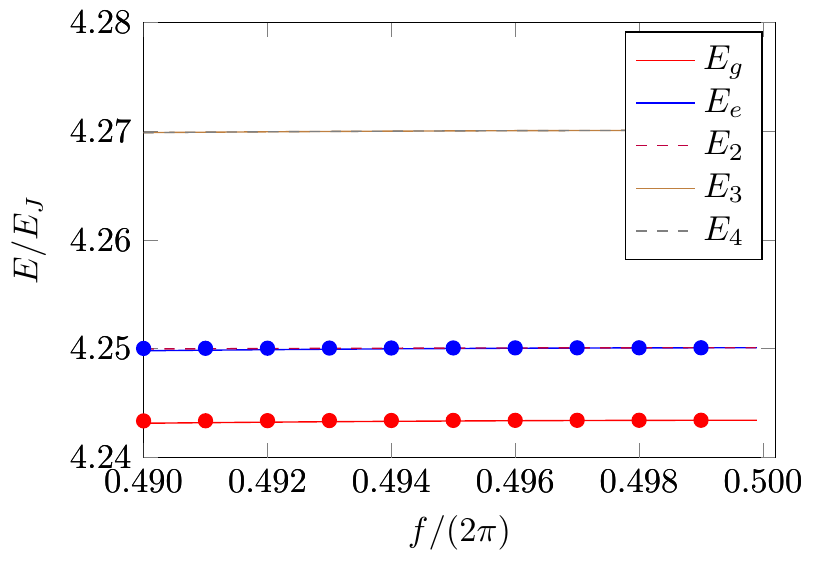}
    \label{fig:Ef}
  } \\
  \subfloat[$g(f)$.]{
    \includegraphics[width=0.4\textwidth]{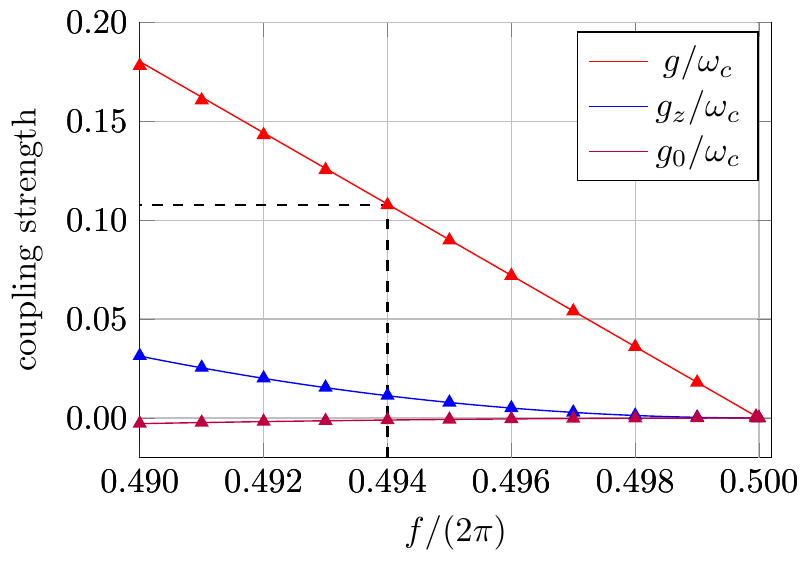}
    \label{fig:gf}
  }
  \subfloat[Zoom in of \protect\subref{fig:Ef}.]{
    \includegraphics[width=0.4\textwidth]{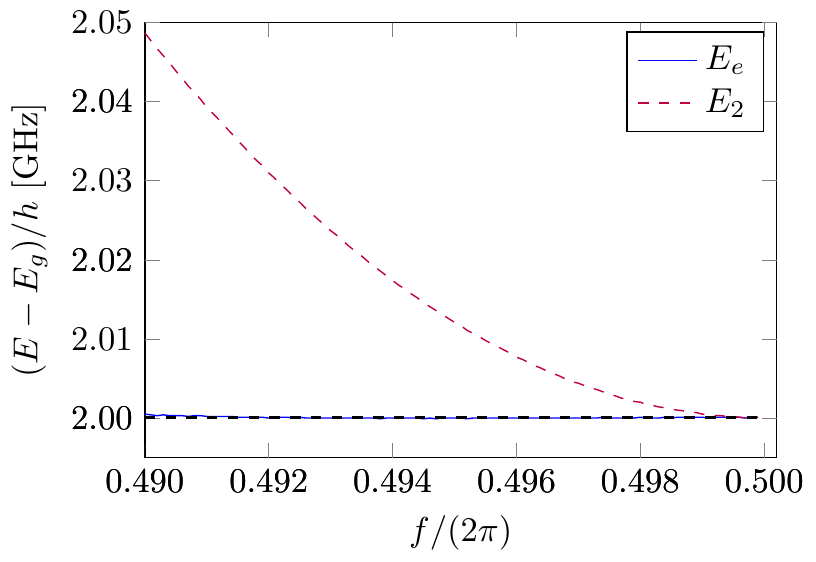}
    \label{fig:DeltaE}
  } \\
  \caption{(Color online). (a) The external flux $f'$ as a function of
  $f$ for a fixed $\delta E$. (b) The five lowest eigenenergies of
  $\hat{H}_{\text{atom}}$ as a function of $f$ for a fixed $\delta E$.
  Results obtained via the numerical diagonalization are represented
  by lines with different colors. $E_{g}$ and $E_{e}$ obtained by
  Eqs.~\eqref{eq:2.B.12} and \eqref{eq:2.B.13} are denoted as red and
  blue solid circles respectively. (c) The coupling strength as a
  function of $f$ for a fixed $\delta E$. The triangles represent
  results obtained via the numerical diagonalization, and the solid
  lines represent results obtained by Eqs.~\eqref{eq:2.B.17},
  \eqref{eq:2.B.18a} and \eqref{eq:2.B.18b}. (d) The energies $E_{e}$
  and $E_{2}$ of two lowest excited states relative to $E_{g}$. The
  black dashed line represents the value of the fixed atomic frequency
  $\delta E/h$. Here we take $\dfrac{E_{J}}{E_{c}} = 150,~E_{J}/h =
  300\rm{GHz}$ and $\delta E/h =\dfrac{\omega_{c}}{2 \pi} =
  2.00005254655 \rm{GHz}$, $L_{0} = 0.06192867473 \rm{nH}$, $L_{r} =
  12.29291953901 \rm{nH}$.}
  \label{fig:tune}
\end{figure*}

From Eqs.~\eqref{eq:2.B.16} and \eqref{eq:2.B.17}, we can get
\begin{equation}
  \begin{split}
    & \dfrac{g^{2} E^{2}_{b}}{8 \omega_{0} \omega_{c} \lambda^{4}
    E^{2}_{J}} = \Delta^{2} e^{-\lambda^{2}} \sin^{2} \dfrac{f'}{2} \\
    = & \dfrac{11 E_{b} \Delta^{2} e^{- \lambda^{2}}}{11 E_{b} + 3
    \lambda^{2} E_{c}} - \dfrac{3 E_{b} E_{c} (\delta E - E_{c})}{(11
    E_{b} + 3 \lambda^{2} E_{c}) E^{2}_{J}}.
  \end{split}
  \label{eq:2.B.19}
\end{equation}
In this article, we attempt to tune the coupling strength $g$ from
zero to a finite value. In order to get an extremely small $g$, we
should have $\delta E > E_{c}$ according to Eq.~\eqref{eq:2.B.19}. In
this sense, $f'$ needs to satisfy the condition that
\begin{equation}
  f' > 2 \pi - \arccos \left(\dfrac{3 \lambda^{4} - 11}{11 + 3
  \lambda^{4}}\right)
  \label{eq:2.B.20}
\end{equation}
for $\delta E$ being fixed.

As illustrated in Fig.~\ref{fig:ff}, we show how to change $f'$ for
different $f$ to keep the atomic energy level splitting $\delta E$
constant. With the same relation between $f^{\prime}$ and $f$, the
five lowest eigenenergies of $\hat{H}_{\text{atom}}$ and the coupling
strength $g$ as functions of $f$ are given in Figs.~\ref{fig:Ef} and
\ref{fig:gf} respectively. When we zoom in Fig.~\ref{fig:Ef}, we
observe that $E_{e} - E_{g}$ is invariant with external flux $f$ which
can seen in Fig.~\ref{fig:DeltaE}.

In Fig.~\ref{fig:tune}\textcolor{red}{c}, we observe that
$g/\omega_{c}$ becomes smaller with the increase of $f$, and it limits
to zero when $f$ tends to $\pi$. We also note that $|g_{0}/g|$ and
$|g_{z}/g|$ increase with the decrease of $f$. When $f$ is far below
$\pi$, $g_{z}$ is comparable with $g$ which makes it nonnegligible. In
this case, our system can simulate a tunable coupling generalized Rabi
model~\cite{PhysRevLett.107.100401}. If we restrict the value range of
$f$ and $f'$, we can get a tunable coupling Rabi model. For example,
as shown in Fig.~\ref{fig:gf}, $|g_{z}/g| \ll 1$ and $|g_{0}/g| \ll 1$
when $f > 0.988 \pi$. Therefore, we can propose the tunable coupling
scheme that changing the values of external flux $f$ and $f'$ to tune
the coupling strength from the extremely weak coupling regime to the
ultrastrong coupling one and leave the energy splitting $\delta E$
unchanged. Moreover, it is worthy to point out that all the results in
Fig.~\ref{fig:tune} obtained by quantum perturbation theory agree well
with those from the numerical exact diagonalization, which verifies
the validness of our calculations.

\section{Dissipation of the qubit\label{sec:3}}

In general, the coupling of a superconducting qubit to its environment
will cause two different dissipative processes, relaxation and pure
dephasing, each with their characteristic time constants $T_{1}$ and
$T_{\varphi}$. In this section, we discuss the performance of the
qubit in terms of these two time constants in the tunable coupling
scheme proposed above.

\subsection{Estimates for the relaxation time $T_{1}$}

As the de-excitation process of a qubit, the relaxation is originated
from a perturbation which couples the qubit with its noise sources. In
this perturbation, the qubit operator is defined as
$\dfrac{\partial \hat{H}_{\rm{atom}}}{\partial \mu}$ where $\mu$ is
the external parameter in the qubit's Hamiltonian which corresponds to
the noise source~\cite{ithier2005decoherence}. For weak noise sources,
we follow Ref.~\cite{schoelkopf2003noise} and give $T^{(\mu)}_{1}$ by
Fermi's golden rule
\begin{equation}
  \Gamma^{(\mu)}_{1} = \dfrac{1}{T^{(\mu)}_{1}} = \dfrac{1}{\hbar^{2}}
  \left|\left\langle e \left| \dfrac{\partial
  \hat{H}_{\text{atom}}}{\partial \mu} \right| g
  \right\rangle\right|^{2} S_{\mu} (\omega),
  \label{eq:3.A.1}
\end{equation}
where $S_{\mu} (\omega)$ is the spectral density of the bath noise. To
this end, we can obtain the relaxation time with Eq.~\eqref{eq:3.A.1}.

Compared to other solid-state qubits~\cite{bouchiat1998quantum,
nakamura1999coherent, friedman2000jr, derwal2000quantum,
koch2007charge, barends2013coherent}, our qubit is remarkably
sensitive to flux noise due to its unique structure. Therefore, it is
reasonable to investigate the qubit's dissipation caused by the flux
noise at first.

Reviewing the qubit architecture in Fig.~\ref{fig:Loop_SQUID}, we find
that the coupling of our qubit to three external magnetic flux biases
opens up an additional channel for energy relaxation, which is the
internal coupling between the circuit and the flux biases through
mutual inductance $M_{i}$ ($i = 1,2,3$). After introducing the flux
noise power spectrum
\begin{equation}
  S_{\Phi_{i}} (\omega) = M^{2}_{i} S_{I_{i}} (\omega) = 2 M^{2}_{i}
  \Theta(\omega) \hbar \omega/Z_{\rm{R}}
  \label{eq:3.A.2}
\end{equation}
with $\omega = \dfrac{\delta E}{\hbar}$ and the environmental
impedance $Z_{\rm{R}}$ for low temperature $k_{B} T \ll \hbar \omega$,
we get
\begin{equation}
  \begin{split}
    T^{(\Phi)}_{1} \approx & \frac{\hbar^{2} \varphi^{2}_{0}
    Z_{\rm{R}}}{E^{2}_{J} \delta E} e^{\lambda^{2}} \left[M^{2}_{1}
    \cos^{2} \frac{\Delta + f'}{2} + M^{2}_{2} \cos^{2} \frac{\Delta -
    f'}{2}\right. \\
    & \left.+ \Delta^{2} M^{2}_{3} \sin^{2} \frac{f'}{2}\right]^{-1}.
  \end{split}
  \label{eq:3.A.3}
\end{equation}
\begin{figure}[!htbp]
  \centering \subfloat[$T^{(\Phi)}_{1} (f)$.]{
    \includegraphics[width=0.4\textwidth, clip, trim = 3 2 2
    1]{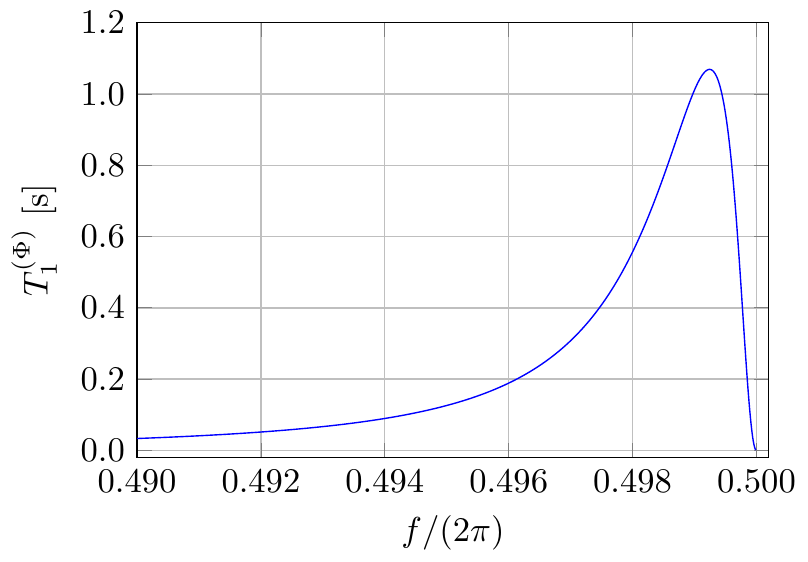}~\label{fig:T1}} \\
    \subfloat[$T^{(\Phi)}_{\varphi} (f)$.]{
    \includegraphics[width=0.4\textwidth, clip, trim = 3 2 2
    1]{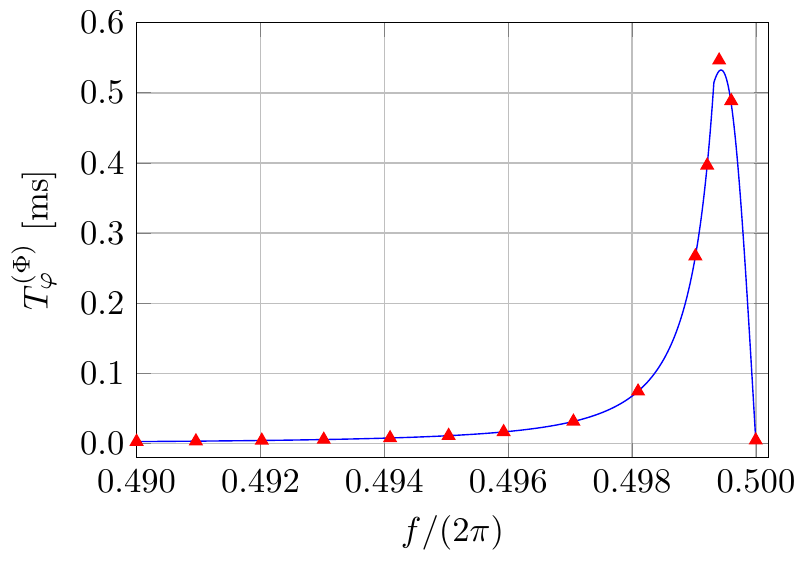}~\label{fig:T2}}
  \caption{(Color online). The characteristic time $T^{(\Phi)}_{1}$
  and $T^{(\Phi)}_{\varphi}$ as a function of the external flux $f$.
  Here we take the same parameters as those of Fig.~\ref{fig:tune} and
  $M_{1} = M_{2} = 40 \Phi_{0}/A$, $M_{3} = 35\Phi_{0}/A$, $Z_{\rm{R}}
  = 50 \Omega$, $A_{\Phi_{1}} = A_{\Phi_{2}} = A_{\Phi_{3}} = 10^{-6}
  \Phi_{0}$. In Fig.~\protect\subref{fig:T2}, the red triangles
  represent results obtained via the numerical diagonalization, and
  the solid lines represent results obtained by Eq.~\eqref{eq:3.B.2}.}
  \label{fig:T}
\end{figure}

With the same qubit for Fig.~\ref{fig:tune} and realistic device
parameters, we plot the relation between the relaxation time
$T^{(\Phi)}_{1}$ and $f$. As shown in Fig.~\ref{fig:T1},
$T^{(\Phi)}_{1}$ reaches its maximum ($1.06893\rm{s}$) near $f =
0.998487\pi$. Obviously, $T^{(\Phi)}_{1}$ is much longer than the
time unit (ns) of a clock cycle used in experiments, which indicates
that the relaxation induced by flux coupling is unlikely to limit the
manipulation of our qubit.

Similar to the flux noise, the charge noise is an important noise
source which limits the applications of charge type superconducting
qubits. To our qubit, the charge noise corresponds to the qubit
operator $\dfrac{\partial \hat{H}_{\rm{atom}}}{\partial \hat{n}_{-}} =
2 E_{c} \hat{n}_{-}$.

Since $\hat{n}_{-} \hat{P} + \hat{P} \hat{n}_{-} = 0$, we have
\begin{equation}
  \langle e | \hat{n}_{-} | g \rangle = \langle e | \hat{n}_{-}
  \hat{P} \hat{P} | g \rangle = - \langle e | \hat{P} \hat{n}_{-}
  \hat{P} | g \rangle = - \langle e | \hat{n}_{-} | g \rangle,
  \label{eq:3.A.4}
\end{equation}
which implies that $\langle e | \hat{n}_{-} | g \rangle = 0$. With
Eqs.~\eqref{eq:3.A.1} and \eqref{eq:3.A.4}, we can show that the
relaxation transition rate of our qubit induced by the charge noise is
zero. In this sense, the charge noise will not affect the relaxation
process of our qubit in the tunable coupling scheme.

\subsection{Estimation of the pure dephasing time
$T_{\varphi}$\label{sec:3B}}

The coupling with the environment results not only in relaxation but
also in pure dephasing. As we know, the origin of dephasing can
be interpreted as the qubit transition frequency fluctuations induced
by noises from outside. In order to study the dephasing of our qubit,
we define $T_{\varphi}$ as the characteristic time for the decay of the
off-diagonal density matrix element. For sufficiently low frequencies,
we assume that the environment provides $1/f$
noise~\cite{paladino20141} to our qubit. In this sense, we
have~\cite{koch2007charge}
\begin{equation}
  T^{(s)}_{\varphi} \simeq \dfrac{\hbar}{A_{s}} \left|\dfrac{\partial
  \delta E}{\partial s}\right|^{-1},
  \label{eq:3.B.1}
\end{equation}
where $A_{s}$ is the $1/f$ amplitude corresponding to the external
parameter represented by $s$ for different noise sources.

According to Eq.~\eqref{eq:2.B.16}, $\delta E$ is dominated by the
external flux and the Josephson energy $E_{J}$, which implies that the
flux noise and the critical current noise are the main noise sources
for our qubit. By Eq.~\eqref{eq:3.B.1}, we have
\begin{equation}
  \begin{split}
    T^{(\Phi)}_{\varphi} = & \dfrac{2 \hbar \varphi_{0}}{\Delta
    A_{\Phi} E^{2}_{J}} e^{\lambda^{2}} \left[\left(\dfrac{11}{3
    E_{c}} - \dfrac{\lambda^{2}}{E_{b}}\right) + \left(\dfrac{11}{3
    E_{c}} + \dfrac{\lambda^{2}}{E_{b}}\right) \right. \\
    & \times \left(\cos f' - \dfrac{3
    \Delta}{2} \sin f'\right) + \left|\left(\dfrac{11}{3 E_{c}} -
    \dfrac{\lambda^{2}}{E_{b}}\right)\right. \\
    & \left.\left. + \left(\dfrac{11}{3 E_{c}} +
    \dfrac{\lambda^{2}}{E_{b}}\right) \left(\cos f' +
    \dfrac{\Delta}{2} \sin f'\right)\right|\right]^{-1}
  \end{split}
  \label{eq:3.B.2}
\end{equation}
with $A_{\Phi} = A_{\Phi_{1}} = A_{\Phi_{2}} = A_{\Phi_{3}}$. For
$A_{\Phi} = 10^{-6} \Phi_{0}$~\cite{yoshihara2006decoherence}, the
same qubit of Fig.~\ref{fig:tune} yields a dephasing time of the order
of $T^{(\Phi)}_{\varphi} \sim 10 \rm{\mu s}$. Fig.~\ref{fig:T2} shows
the variation of $T^{(\Phi)}_{\varphi}$ with $f$ in our tunable
coupling scheme.

For the critical current noise, we choose
$A_{I_{c}} = 10^{-6} I_{c}$~\cite{van2004decoherence}. Then
Eq.~\eqref{eq:3.B.1} gives
\begin{equation}
  T^{(I_{c})}_{\varphi} \simeq \dfrac{\hbar}{A_{I_{c}}}
  \left|\dfrac{\partial \delta E}{\partial I_{c}}\right|^{-1} =
  \dfrac{\hbar}{2 (A_{I_{c}}/I_{c}) (\delta E - E_{c})} \approx
  1.51442 s
  \label{eq:3.B.3}
\end{equation}
for our qubit in Fig.~\ref{fig:tune}.

Furthermore, the fluctuation of our qubit transition frequency can
also be induced by the fluctuation of the relative cooper pair number
between the two dc-SQUIDs. Assuming that the relative cooper pair number
$\hat{n}_{c}$ can be decomposed into the noiseless charge number
$\hat{n}_{-}$ and a small noise term, i.e., $\hat{n}_{c} = \hat{n}_{-}
+ \delta n_{-}$ with $\delta n_{-} \ll n_{-}$. Then a Taylor expansion
of $\hat{H}_{\rm{atom}}$ yields
\begin{equation}
  \hat{H}_{\rm{atom}} \to \hat{H}_{0} + \hat{V} + 2 E_{c}
  \hat{n}_{-} \delta n_{-}.
  \label{eq:3.B.4}
\end{equation}

Considering the relation between $\hat{n}_{-}$ and the parity operator
$\hat{P}$, we introduce the odd-parity excited state $| \psi_{-}
\rangle$ with its corresponding eigenenergy
\begin{equation}
  E_{2} = \dfrac{E_{b}}{2} + E_{c} + 4 E_{J} - \Delta^{2} E^{2}_{J}
  e^{- \lambda^{2}} \left(\dfrac{\cos^{2} \dfrac{f'}{2}}{3 E_{c}} +
  \dfrac{\lambda^{2} \sin^{2} \dfrac{f'}{2}}{E_{b} + 3 E_{c}}\right).
  \label{eq:3.B.5}
\end{equation}
Therefore, we use
\begin{subequations}
  \begin{align}
    \langle \psi_{-} | \hat{n}_{-} | g \rangle & \approx - \dfrac{i
    \sqrt{2} \Delta E_{J} e^{- \lambda^{2}/2} \cos
    \dfrac{f'}{2}}{E_{c}}, \label{eq:3.B.6a} \\
    \langle \psi_{-} | \hat{n}_{-} | e \rangle & \approx 1 + 2
    \Delta^{2} E^{2}_{J} e^{- \lambda^{2}} \left(\dfrac{\cos^{2}
    \dfrac{f'}{2}}{9 E^{2}_{c}} - \dfrac{\lambda^{2} \sin^{2}
    \dfrac{f'}{2}}{(E_{b} + 3 E_{c})^{2}}\right) \label{eq:3.B.6b}
  \end{align}
  \label{eq:3.B.6}
\end{subequations}
to get the modification of the energy level splitting $\delta E
(n_{-})$ from the coupling between $| \psi_{-} \rangle$ and $| e
\rangle$, $| g \rangle$ by the charge noise term $2 E_{c} \hat{n}_{-}
\delta n_{-}$
\begin{equation}
  \begin{split}
    \delta [\delta E] = & \delta E (n_{-} + \delta n_{-}) - \delta E
    (n_{-}) \\
    = & 4 E^{2}_{c} (\delta n_{-})^{2} \left(\dfrac{|\langle \psi_{-}
    | \hat{n}_{-} | g \rangle|^{2}}{E_{2} - E_{g}} - \dfrac{|\langle
    \psi_{-} | \hat{n}_{-} | e \rangle|^{2}}{E_{2} - E_{e}}\right),
  \end{split}
  \label{eq:3.B.7}
\end{equation}
which is proportional to the square of $\delta n_{-}$. In other words,
the second order contributions of the charge noise will dominate the
pure dephasing process.
\begin{figure}[!htbp]
  \centering \subfloat[$T^{(c)}_{\varphi} (f)$.]{
    \includegraphics[width=0.4\textwidth, clip, trim = 3 2 2
    1]{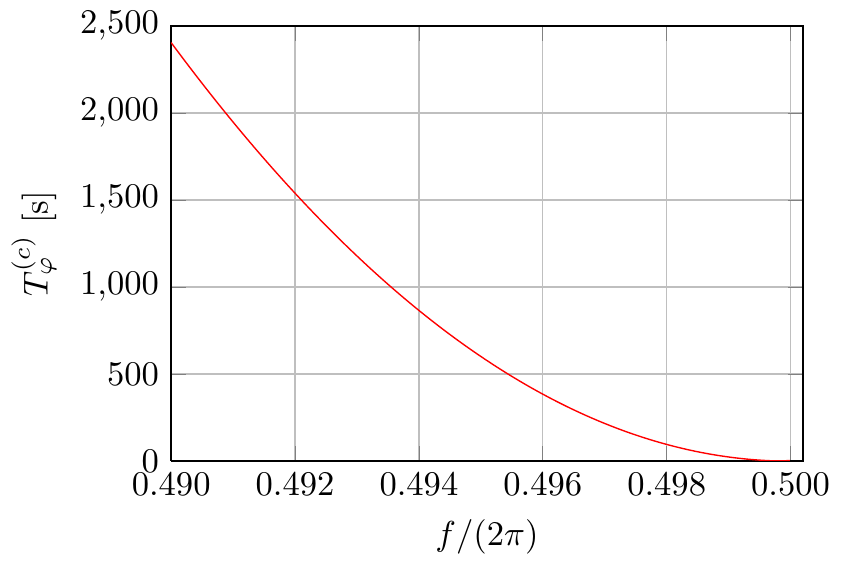}~\label{fig:T2CN}} \\
    \subfloat[Zoom in of \protect\subref{fig:T2CN}.]{
    \includegraphics[width=0.4\textwidth, clip, trim = 3 2 2
    1]{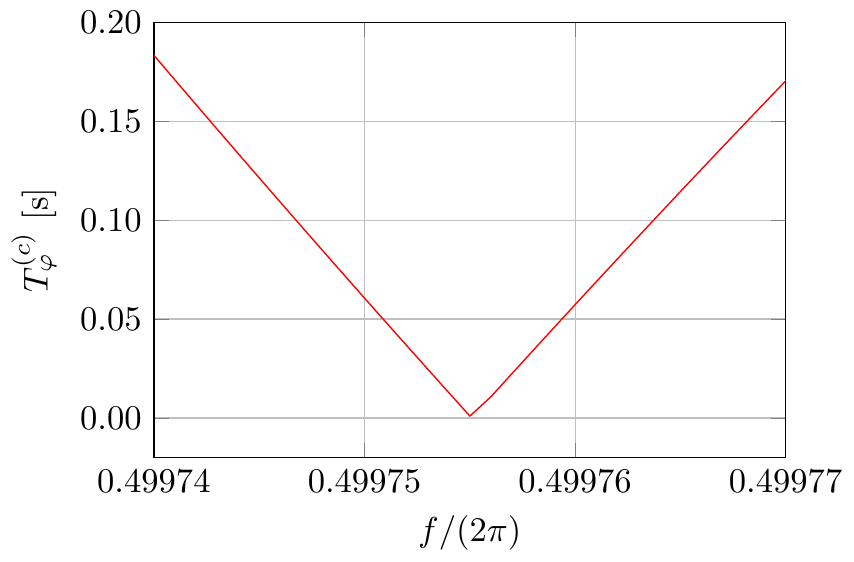}~\label{fig:T2CN_zoom}}
  \caption{(Color online). The characteristic time $T^{(c)}_{\varphi}$
  as a function of the external flux $f$. Here we take the same
  parameters as those of Fig.~\ref{fig:tune} and $A_{c} = 10^{-4} e$.}
  \label{fig:T2c}
\end{figure}

Based on the above consideration, we generalize Eq.~\eqref{eq:3.B.1}
to the second order
\begin{equation}
  \begin{split}
    T^{(c)}_{\varphi} \simeq & \left|\dfrac{\pi^{2} A^{2}_{c}}{\hbar}
    \dfrac{\partial^{2} \delta E}{\partial (e \delta
    n_{-})^{2}}\right|^{-1} \\
    = & \dfrac{\hbar e^{2}}{\pi^{2} A^{2}_{c}} \left|\lim_{\delta
    n_{-} \to 0} \left[\dfrac{\delta E (n_{-} + 2 \delta n_{-}) -
    \delta E (n_{-} + \delta n_{-})}{(\delta n_{-})^{2}}\right.\right.
    \\
    & \quad\qquad\left.\left. - \dfrac{\delta E (n_{-} + \delta n_{-})
    - \delta E (n_{-})}{(\delta n_{-})^{2}}\right]\right|^{-1} \\
    = & \dfrac{\hbar e^{2}}{8 \pi^{2} A^{2}_{c} E^{2}_{c}}
    \left|\dfrac{|\langle \psi_{-} | \hat{n}_{-} | g
    \rangle|^{2}}{E_{2} - E_{g}} - \dfrac{|\langle \psi_{-} |
    \hat{n}_{-} | e \rangle|^{2}}{E_{2} - E_{e}}\right|^{-1},
  \end{split}
  \label{eq:3.B.8}
\end{equation}
where $A_{c}$ is the amplitude of the charge $1/f$ noise.

With the same parameters in Fig.~\ref{fig:tune} and $A_{c} =
10^{-4} e$~\cite{PhysRevB.53.13682}, we find that $T^{(c)}_{\varphi}$
reaches its minimum $1.00293\rm{ms}$ near $f = 0.99951\pi$ which can
be seen in Fig.~\ref{fig:T2c}. In this sense, our qubit has an
excellent performance suppressing the charge noise.

\section{Estimation of device parameters\label{sec:4}}

In order to apply our design to experiments, we should choose proper
value of device parameters such as $L_{0}$, $L_{r}$. In this paper, we
attempt to tune the atom-photon coupling strength from weak coupling
regime into the ultra strong one while keeping the cavity frequency
and the atomic frequency invariant. With Eq.~\eqref{eq:2.B.17}, we can
write the formula of the coupling strength with respect to the cavity
frequency
\begin{equation}
  \dfrac{g}{\omega_{c}} = \sqrt{\dfrac{8 \omega_{0}}{\omega_{c}}}
  \Delta \lambda^{4} \dfrac{E_{J}}{E_{c}} e^{- \lambda^{2}/2} \sin
  \dfrac{f'}{2},
  \label{eq:4.1}
\end{equation}
which indicates that $\lambda$, the ratio $\dfrac{E_{J}}{E_{c}}$ and
$\dfrac{\omega_{0}}{\omega_{c}}$ are the main factors influencing its
value.

During the derivation process of $\delta E$ and $g$, we assume that
$\Delta \ll \pi$ and $\lambda \ll 1$ ($E_{b} \gg E_{c}$) to ensure the
establishment of quantum perturbation theory. In this sense, we have
\begin{equation}
  \dfrac{1}{2 L_{0}} + \dfrac{1}{L_{r}} = \dfrac{1}{L'_{r}} \gg
  \dfrac{E_{c}}{8 \varphi^{2}_{0}},
  \label{eq:4.2}
\end{equation}
which means that $L_{0}$ and $L_{r}$ are both far more less than
$\dfrac{8 \varphi^{2}_{0}}{E_{c}}$. In Fig.~\ref{fig:tune}, we take
$E_{c}/h = 2 \rm{GHz}$, then Eq.~\eqref{eq:4.2} gives $\{L_{0},
L_{r}\} \ll 0.653983 \rm{\mu H}$ which implies the reasonableness of
our choice.

Generally, the relaxation time of a qubit in a circuit QED experiment
should be long enough. According to Eqs.~\eqref{eq:3.A.3},
\eqref{eq:3.B.2}, \eqref{eq:3.B.3} and \eqref{eq:3.B.8}, the
characteristic time $T_{1}$ and $T_{\varphi}$ are proportional to the
ratio $\dfrac{E_{J}}{E_{c}}$ once $E_{J}$ is fixed. It seems that we
need to take $\dfrac{E_{J}}{E_{c}} \gg 1$ to get a sufficiently large
coupling strength. However, as expressed in Eqs.~\eqref{eq:2.B.18a}
and \eqref{eq:2.B.18b}, the ratio $\dfrac{g_{z}}{g}$ and
$\dfrac{g_{0}}{g}$ are inversely proportional to
$\dfrac{E_{J}}{E_{c}}$. Therefore the value of $\dfrac{E_{J}}{E_{c}}$
should be restricted by $\left|\dfrac{g_{z}}{g}\right| \ll 1$ if we
want to simulate the Rabi model.

Besides, focusing on Eq.~\eqref{eq:4.1}, we can find that the
transverse coupling strength is proportional to the square root of
$\dfrac{\omega_{0}}{\omega_{c}}$. This indicates that we need to take
a sufficiently small $L_{0}$ to obtain a relatively large
$\dfrac{g}{\omega_{c}}$.

\section{Conclusion\label{sec:5}}

In this article, we present a theoretical proposal with a tunable
coupling between an artificial two-level atom and a waveguide
transmission line resonator by controlling the external magnetic
fluxes. In our scheme, the coupling can be continuously tuned from
zero to the ultrastrong regime while keeping fixed atomic level
splitting. We also investigate the performance of our qubit under the
influences of the environment, and find that our system operates well
against the main noises. Our analytical results are based on quantum
perturbation theory with the parity symmetry, which are verified by
the numerical simulations. In order to apply our qubit design to
experiments, we discuss how to choose the device parameters by our
analytical results. We hope that our work will stimulate the coherent
manipulation of the circuit QED system in the fields of quantum
simulation and quantum computing.

\begin{acknowledgments}
  This work is supported by NSF of China (Grant Nos. 11475254
  and 11775300), NKBRSF of China (Grant No. 2014CB921202), the
  National Key Research and Development Program of China
  (2016YFA0300603).
\end{acknowledgments}

\clearpage

\appendix
\begin{widetext}
\section{Derivation of the energy splitting and the coupling
strength\label{app:A}}

Because $E_{b} \gg E_{c}$, we can restrict ourselves in the subspace
with the bases $\{|n=0\rangle, |n=1\rangle\}$. In this subspace, the
term
\begin{equation}
  \begin{split}
  & \cos(\phi_{+} + \frac{f^{\prime}}{2}) = \cos \frac{f^{\prime}}{2}
  \cos \phi_{+} - \sin \frac{f^{\prime}}{2} \sin \phi_{+} \\
  = & \cos \frac{f^{\prime}}{2} \left(\langle 0|\cos \phi_{+}|0\rangle
  |0\rangle \langle 0| + \langle 1|\cos \phi_{+}|1\rangle |1\rangle
  \langle 1| \right) - \sin \frac{f^{\prime}}{2} \left(\langle 0|\sin
  \phi_{+} |1\rangle |0\rangle \langle 1| + \langle 1|\sin \phi_{+}
  |0\rangle |1\rangle \langle 0| \right) \\
  = & \cos \frac{f^{\prime}}{2} \left(e^{-\lambda^{2}/2} |0\rangle
  \langle 0| + (1-\lambda^{2}) e^{-\lambda^{2}/2} |1\rangle \langle 1|
  \right) - \sin \frac{f^{\prime}}{2} \lambda e^{-\lambda^{2}/2}
  \left(|0\rangle \langle 1| +  |1\rangle \langle 0| \right),
  \end{split}
  \label{eq:app:A.1}
\end{equation}
where we use the following expressions:
\begin{equation}
  \langle 0 | \cos \phi_{+} | 0 \rangle = \frac{1}{2}
  \left(\langle 0 | e^{i \lambda (\hat{b} + \hat{b}^{\dagger})} | 0
  \rangle + \langle 0 | e^{-i \lambda (\hat{b} + \hat{b}^{\dagger})}
  | 0 \rangle\right) = \frac{e^{- \lambda^{2}/2}}{2} \left( \langle 0
  | e^{i \lambda \hat{b}^{\dagger}} e^{i \lambda \hat{b}} | 0 \rangle
  + \langle 0 | e^{-i \lambda \hat{b}^{\dagger}} e^{-i \lambda
  \hat{b}} | 0 \rangle\right) = e^{- \lambda^{2}/2},
  \label{eq:app:A.2}
\end{equation}
\begin{equation}
  \langle 1|\cos \phi_{+}|1\rangle = \frac{1}{2} \left(\langle 1 |
  e^{i \lambda (\hat{b} + \hat{b}^{\dagger})} | 1 \rangle + \langle
  1 | e^{-i \lambda (\hat{b} + \hat{b}^{\dagger})} | 1
  \rangle\right) = \frac{e^{- \lambda^{2}/2}}{2} \left(\langle 1 |
  e^{i \lambda \hat{b}^{\dagger}} e^{i \lambda \hat{b}} | 1 \rangle +
  \langle 1 | e^{-i \lambda \hat{b}^{\dagger}} e^{-i \lambda \hat{b}}
  | 1 \rangle\right) = (1-\lambda^{2}) e^{- \lambda^{2}/2},
  \label{eq:app:A.3}
\end{equation}
\begin{equation}
  \langle 1 | \sin \phi_{+} | 0 \rangle = \frac{1}{2i}
  \left(\langle 1 | e^{i \lambda (\hat{b} + \hat{b}^{\dagger})} | 0
  \rangle - \langle 1 |e^{-i \lambda (\hat{b} + \hat{b}^{\dagger})}
  | 0 \rangle\right) = \frac{e^{- \lambda^{2}/2}}{2i} \left(\langle 1
  | e^{i \lambda \hat{b}^{\dagger}} e^{i \lambda \hat{b}} | 0 \rangle
  - \langle 1 | e^{-i \lambda \hat{b}^{\dagger}} e^{-i \lambda
  \hat{b}} | 0 \rangle\right) = \lambda e^{- \lambda^{2}/2},
  \label{eq:app:A.4}
\end{equation}
\begin{equation}
  \langle 0 | \sin \phi_{+} | 1 \rangle = (\langle 1 | \sin \phi_{+} |
  0 \rangle)^{\ast}  = \lambda e^{- \lambda^{2}/2}.
  \label{eq:app:A.5}
\end{equation}

In addition, in the subspace with even parity $\{ | 0 \rangle, | +
\rangle_{n}\}$, the term
\begin{equation}
  -2 \sin \left(\phi_{-} + \frac{\Delta}{2}\right) = i \left(e^{i
  \left(\phi_{-} + \frac{\Delta}{2}\right)} - e^{- i
  \left(\phi_{-} + \frac{\Delta}{2}\right)}\right) = i \sum_{n>0}
  \left(| +_{n+1} \rangle \langle +_{n} | - | +_{n}
  \rangle \langle +_{n+1} |\right) + i \sqrt{2} (| +_{1} \rangle
  \langle 0 | - | 0 \rangle \langle +_{1} |).
  \label{eq:app:A.6}
\end{equation}

Thus we obtain the expression for the perturbation $\hat{V}$ in the
relative subspace for our problem. Using the perturbation theory, we
obtain
\begin{equation}
  \begin{split}
    | g \rangle = & | 0; 0 \rangle - \dfrac{\langle 0; +_{1} | \hat{V}
    | 0; 0 \rangle}{E_{c}} | 0; +_{1} \rangle - \dfrac{\langle 1;
    +_{1} | \hat{V} | 0; 0 \rangle}{E_{b} + E_{c}} | 1; +_{1} \rangle
    + \dfrac{\langle 0; +_{2} | \hat{V} | 0; +_{1} \rangle \langle
    0; +_{1} | \hat{V} | 0; 0 \rangle}{4 E^{2}_{c}} | 0; +_{2} \rangle
    \\
    & + \dfrac{\langle 0; +_{2} | \hat{V} | 1; +_{1} \rangle \langle
    1; +_{1} | \hat{V} | 0; 0 \rangle}{4 E_{c} (E_{b} + E_{c})} | 0;
    +_{2} \rangle + \dfrac{\langle 1; +_{2} | \hat{V} | 0; +_{1}
    \rangle \langle 0; +_{1} | \hat{V} | 0; 0 \rangle}{(E_{b} + 4
    E_{c}) E_{c}} | 1; +_{2} \rangle \\
    & + \dfrac{\langle 1; +_{2} | \hat{V} | 1; +_{1} \rangle \langle
    1; +_{1} | \hat{V} | 0; 0 \rangle}{(E_{b} + 4 E_{c}) (E_{b} +
    E_{c})} | 1; +_{2} \rangle + \dfrac{\langle 1; 0 | \hat{V} | 0;
    +_{1} \rangle \langle 0; +_{1} | \hat{V} | 0; 0 \rangle}{E_{b}
    E_{c}} | 1; 0 \rangle \\
    & + \dfrac{\langle 1; 0 | \hat{V} | 1; +_{1} \rangle \langle 1;
    +_{1} | \hat{V} | 0; 0 \rangle}{E_{b} (E_{b} + E_{c})} | 1; 0
    \rangle, \\
  \end{split}
  \label{eq:app:A.7}
\end{equation}
and
\begin{equation}
  \begin{split}
    | e \rangle = & | 0; +_{1} \rangle - \dfrac{\langle 0; +_{2} |
    \hat{V} | 0; +_{1} \rangle}{3 E_{c}} | 0; +_{2} \rangle +
    \dfrac{\langle 0; 0 | \hat{V} | 0; +_{1} \rangle}{E_{c}} | 0; 0
    \rangle - \dfrac{\langle 1; +_{2} | \hat{V} | 0; +_{1}
    \rangle}{E_{b} + 3 E_{c}} | 1; +_{2} \rangle - \dfrac{\langle 1; 0
    | \hat{V} | 0; +_{1} \rangle}{E_{b} - E_{c}} | 1; 0 \rangle \\
    & + \dfrac{\langle 0; +_{3} | \hat{V} | 0; +_{2} \rangle \langle
    0; +_{2} | \hat{V} | 0; +_{1} \rangle}{24 E^{2}_{c}} | 0; +_{3}
    \rangle + \dfrac{\langle 0; +_{3} | \hat{V} | 1; +_{2} \rangle
    \langle 1; +_{2} | \hat{V} | 0; +_{1} \rangle}{8 E_{c} (E_{b} + 3
    E_{c})} | 0; +_{3} \rangle \\
    & + \dfrac{\langle 1; +_{3} | \hat{V} | 0; +_{2} \rangle \langle
    0; +_{2} | \hat{V} | 0; +_{1} \rangle}{3 E_{c} (E_{b} + 8 E_{c})}
    | 1; +_{3} \rangle + \dfrac{\langle 1; +_{3} | \hat{V} | 1; +_{2}
    \rangle \langle 1; +_{2} | \hat{V} | 0; +_{1} \rangle}{(E_{b} + 8
    E_{c}) (E_{b} + 3 E_{c})} | 1; +_{3} \rangle \\
    & + \dfrac{\langle 1; +_{1} | \hat{V} | 0; +_{2} \rangle \langle
    0; +_{2} | \hat{V} | 0; +_{1} \rangle}{3 E_{b} E_{c}} | 1; +_{1}
    \rangle - \dfrac{\langle 1; +_{1} | \hat{V} | 0; 0 \rangle \langle
    0; 0 | \hat{V} | 0; +_{1} \rangle}{E_{b} E_{c}} | 1; +_{1} \rangle
    \\
    & + \dfrac{\langle 1; +_{1} | \hat{V} | 1; +_{2} \rangle \langle
    1; +_{2} | \hat{V} | 0; +_{1} \rangle}{E_{b} (E_{b} + 3 E_{c})} |
    1; +_{1} \rangle + \dfrac{\langle 1; +_{1} | \hat{V} | 1; 0
    \rangle \langle 1; 0 | \hat{V} | 0; +_{1} \rangle}{E_{b} (E_{b} -
    E_{c})} | 1; +_{1} \rangle,
  \end{split}
  \label{eq:app:A.8}
\end{equation}
where
\begin{subequations}
  \begin{align}
    & \langle 0; +_{1} | \hat{V} | 0; 0 \rangle = i \sqrt{2} \Delta
    E_{J} e^{- \lambda^{2}/2} \cos \dfrac{f'}{2}, \label{eq:app:A.9a}
    \\
    & \langle 1; +_{1} | \hat{V} | 0; 0 \rangle = \langle 0; +_{1} |
    \hat{V} | 1; 0 \rangle = - i \sqrt{2} \Delta E_{J} \lambda e^{-
    \lambda^{2}/2} \sin \dfrac{f'}{2}, \label{eq:app:A.9b} \\
    & \langle 0; +_{2} | \hat{V} | 0; +_{1} \rangle = \langle 0; +_{3}
    | \hat{V} | 0; +_{2} \rangle = i \Delta E_{J} e^{- \lambda^{2}/2}
    \cos \dfrac{f'}{2}, \label{eq:app:A.9c} \\
    & \langle 1; +_{2} | \hat{V} | 0; +_{1} \rangle = \langle 0; +_{2}
    | \hat{V} | 1; +_{1} \rangle = \langle 0; +_{3} | \hat{V} | 1;
    +_{2} \rangle = \langle 1; +_{3} | \hat{V} | 0; +_{2} \rangle = -i
    \Delta E_{J} \lambda e^{- \lambda^{2}/2} \sin \dfrac{f'}{2},
    \label{eq:app:A.9d} \\
    & \langle 1; +_{2} | \hat{V} | 1; +_{1} \rangle = \langle 1; +_{3}
    | \hat{V} | 1; +_{2} \rangle = i \Delta E_{J} (1 - \lambda^{2})
    e^{- \lambda^{2}/2} \cos \dfrac{f'}{2}, \label{eq:app:A.9e} \\
    & \langle 1; 0 | \hat{V} | 1; +_{1} \rangle = -i \sqrt{2} \Delta
    E_{J} (1 - \lambda^{2}) e^{- \lambda^{2}/2} \cos \dfrac{f'}{2}.
    \label{eq:app:A.9f}
  \end{align}
  \label{eq:app:A.9}
\end{subequations}

Therefore
\begin{equation}
  \begin{split}
    E_{g} = & \frac{1}{2} E_{b} + 4 E_{J} - 2 \Delta^{2}
    \frac{E_{J}^{2}}{E_{c}} \cos^{2} \frac{f^{\prime}}{2} e^{-
    \lambda^{2}} - 2 \Delta^{2} \frac{E_{J}^{2}}{E_{b} + E_{c}}
    \sin^{2} \frac{f^{\prime}}{2} \lambda^{2} e^{-\lambda^{2}} \\
    \approx & \frac{1}{2} E_{b} + 4 E_{J} - 2 \Delta^{2} E_{J}^{2}
    e^{-\lambda^{2}} \left(\dfrac{\cos^{2}
    \dfrac{f^{\prime}}{2}}{E_{c}} + \dfrac{\lambda^{2} \sin^{2}
    \dfrac{f^{\prime}}{2}}{E_{b}}\right),
  \end{split}
  \label{eq:app:A.10}
\end{equation}
\begin{equation}
  \begin{split}
  E_{e} = & \frac{1}{2} E_{b} + E_{c} + 4 E_{J} + \frac{5}{3}
  \Delta^{2} \frac{E_{J}^{2}}{E_{c}} \cos^{2} \frac{f^{\prime}}{2}
  e^{-\lambda^{2}} - 2 \Delta^{2} \frac{E_{J}^{2}}{E_{b} - E_{c}}
  \sin^{2} \frac{f^{\prime}}{2} \lambda^{2} e^{-\lambda^{2}} -
  \Delta^{2} \frac{E_{J}^{2}}{E_{b} + 3 E_{c}} \sin^{2}
  \frac{f^{\prime}}{2} \lambda^{2} e^{-\lambda^{2}} \\
  \approx & \frac{1}{2} E_{b} + E_{c} + 4 E_{J} + \Delta^{2} E_{J}^{2}
  e^{-\lambda^{2}} \left(\dfrac{5 \cos^{2} \dfrac{f^{\prime}}{2}}{3
  E_{c}} - \dfrac{3 \lambda^{2} \sin^{2} \dfrac{f^{\prime}}{2
  }}{E_{b}}\right).
  \end{split}
  \label{eq:app:A.11}
\end{equation}

Then the energy splitting is
\begin{equation}
  \begin{split}
  \delta E = & E_{e} - E_{g} = E_{c} + \frac{11}{3} \Delta^{2}
  \frac{E_{J}^{2}}{E_{c}} \cos^{2} \frac{f^{\prime}}{2}
  e^{-\lambda^{2}} - 4 \Delta^{2} \frac{E_{c} E_{J}^{2}}{E_{b}^{2} -
  E_{c}^{2}} \sin^{2} \frac{f^{\prime}}{2} \lambda^{2}
  e^{-\lambda^{2}} -  \Delta^{2} \frac{E_{J}^{2}}{E_{b} + 3 E_{c}}
  \sin^{2} \frac{f^{\prime}}{2} \lambda^{2} e^{-\lambda^{2}} \\
  \approx & E_{c} + \Delta^{2} E_{J}^{2} e^{-\lambda^{2}}
  \left(\frac{11}{3 E_{c}} \cos^{2} \frac{f^{\prime}}{2} -
  \dfrac{\lambda^{2}}{E_{b}} \sin^{2} \dfrac{f'}{2}\right).
  \end{split}
  \label{eq:app:A.12}
\end{equation}
The approximation is valid due to the assumption $E_{b} \gg E_{c}$.

Now we can calculate the operator $\hat{\phi}_{+}$ up to the second
order:
\begin{equation}
  \begin{split}
    \langle g | \hat{\phi}_{+} | g \rangle \approx & \lambda \left(
    \dfrac{\langle 0; 0 | \hat{V} | 0; +_{1} \rangle \langle 1; +_{1}
    | \hat{V} | 0; 0 \rangle}{E_{c} (E_{b} + E_{c})} + \dfrac{\langle
    1; 0 | \hat{V} | 0; +_{1} \rangle \langle 0; +_{1} | \hat{V} | 0;
    0 \rangle}{E_{b} E_{c}} + \dfrac{\langle 1; 0 | \hat{V} | 1; +_{1}
    \rangle \langle 1; +_{1} | \hat{V} | 0; 0 \rangle}{E_{b} (E_{b} +
    E_{c})} + \text{h.c.}\right) \\
    = & - 2 \Delta^{2} E^{2}_{J} \lambda^{2} e^{- \lambda^{2}} \sin f'
    \left[\dfrac{1}{E_{c} (E_{b} + E_{c})} + \dfrac{1}{E_{b} E_{c}} +
    \dfrac{1 - \lambda^{2}}{E_{b} (E_{b} + E_{c})}\right] \\
    = & - 2 \Delta^{2} E^{2}_{J} \lambda^{2} \dfrac{2 E_{b} + (2 -
    \lambda^{2}) E_{c}}{E_{b} E_{c} (E_{b} + E_{c})} e^{- \lambda^{2}}
    \sin f' \approx - \dfrac{4 \Delta^{2} E^{2}_{J} \lambda^{2} e^{-
    \lambda^{2}} \sin f'}{E_{b} E_{c}}, \\
  \end{split}
  \label{eq:app:A.13}
\end{equation}
\begin{equation}
  \begin{split}
    \langle e | \hat{\phi}_{+} | e \rangle \approx & \lambda
    \left(\dfrac{\langle 1; +_{2} | \hat{V} | 0; +_{1} \rangle \langle
    0; +_{1} | \hat{V} | 0; +_{2} \rangle}{3 E_{c} (E_{b} + 3
    E_{c})} - \dfrac{\langle 1; 0 | \hat{V} | 0; +_{1} \rangle \langle
    0; +_{1} | \hat{V} | 0; 0 \rangle}{E_{c} (E_{b} - E_{c})} +
    \dfrac{\langle 1; +_{1} | \hat{V} | 0; +_{2} \rangle \langle
    0; +_{2} | \hat{V} | 0; +_{1} \rangle}{3 E_{b} E_{c}}\right. \\
    & \left.- \dfrac{\langle 1; +_{1} | \hat{V} | 0; 0 \rangle \langle
    0; 0 | \hat{V} | 0; +_{1} \rangle}{E_{b} E_{c}} + \dfrac{\langle
    1; +_{1} | \hat{V} | 1; +_{2} \rangle \langle 1; +_{2} | \hat{V} |
    0; +_{1} \rangle}{E_{b} (E_{b} + 3 E_{c})} + \dfrac{\langle 1;
    +_{1} | \hat{V} | 1; 0 \rangle \langle 1; 0 | \hat{V} | 0; +_{1}
    \rangle}{E_{b} (E_{b} - E_{c})} + \text{h.c.}\right) \\
    = & 2 \Delta^{2} E^{2}_{J} \lambda^{2} e^{- \lambda^{2}}
    \left[\sin f' \left(\dfrac{5}{6 E_{b} E_{c}} + \dfrac{1}{E_{c}
    (E_{b} - E_{c})} - \dfrac{3 (1 - \lambda^{2}) E_{c} + E_{b}}{6
    E_{b} E_{c} (E_{b} + 3 E_{c})}\right) - \dfrac{\lambda (1 - \cos
    f')}{E_{b} (E_{b} - E_{c})}\right] \\
    = & 2 \Delta^{2} E^{2}_{J} \lambda^{2} e^{- \lambda^{2}}
    \left[\dfrac{\lambda (\cos f' - 1)}{E_{b} (E_{b} - E_{c})} +
    \dfrac{10 E^{2}_{b} + (26 + 3 \lambda^{2}) E_{b} E_{c} - 3 (4 +
    \lambda^{2}) E^{2}_{c}}{6 E_{b} E_{c} (E_{b} + 3 E_{c}) (E_{b} -
    E_{c})} \sin f'\right] \\
    \approx & \dfrac{10}{3} \Delta^{2} \dfrac{E^{2}_{J}}{E_{b} E_{c}}
    \lambda^{2} e^{- \lambda^{2}} \sin f',
  \end{split}
  \label{eq:app:A.14}
\end{equation}
and
\begin{equation}
  \langle e | \hat{\phi}_{+} | g \rangle \approx - \dfrac{\lambda
  \langle 1; +_{1} | \hat{V} | 0; 0 \rangle}{E_{b} + E_{c}} -
  \dfrac{\lambda (\langle 1; 0 | \hat{V} | 0; +_{1}
  \rangle)^{\ast}}{E_{b} - E_{c}} = \dfrac{i 2 \sqrt{2} \Delta E_{J}
  E_{b} \lambda^{2} e^{-\lambda^{2}/2}}{E^{2}_{b} - E^{2}_{c}} \sin
  \dfrac{f'}{2} \approx \dfrac{i 2 \sqrt{2} \Delta E_{J} \lambda^{2}
  e^{-\lambda^{2}/2}}{E_{b}} \sin \dfrac{f'}{2}.
  \label{eq:app:A.15}
\end{equation}

With Eqs.~\eqref{eq:app:A.13}, \eqref{eq:app:A.14} and
\eqref{eq:app:A.15}, we can rewrite Eq.~\eqref{eq:2.B.11b} as
\begin{equation}
  \hat{H}_{\text{int}} = \hbar (g_{x} \hat{\sigma}_{x} + g
  \hat{\sigma}_{y} + g_{0} \hat{\sigma}_{0} + g_{z} \hat{\sigma}_{z})
  (\hat{a}^{\dagger} + \hat{a}),
  \label{eq:app:A.16}
\end{equation}
where
\begin{subequations}
  \begin{align}
    g_{x} = & - \sqrt{\omega_{0} \omega_{c}} \dfrac{\langle e |
    \hat{\phi}_{+} | g \rangle + \langle g | \hat{\phi}_{+} | e
    \rangle}{2} = 0, \label{eq:app:A.17a} \\
    g = & -i \sqrt{\omega_{0} \omega_{c}} \dfrac{\langle e |
    \hat{\phi}_{+} | g \rangle - \langle g | \hat{\sigma}_{+} | e
    \rangle}{2} \approx \dfrac{\sqrt{8 \omega_{0} \omega_{c}} \Delta
    E_{J} \lambda^{2} e^{- \lambda^{2}/2}}{E_{b}} \sin \dfrac{f'}{2},
    \label{eq:app:A.17b} \\
    g_{0} = & - \sqrt{\omega_{0} \omega_{c}} \dfrac{\langle e |
    \hat{\phi}_{+} | e \rangle + \langle g | \hat{\phi}_{+} | g
    \rangle}{2} \approx \dfrac{\sqrt{\omega_{0} \omega_{c}}
    \Delta^{2} E^{2}_{J} \lambda^{2} e^{- \lambda^{2}}}{3 E_{b} E_{c}}
    \sin f', \label{eq:app:A.17c} \\
    g_{z} = & - \sqrt{\omega_{0} \omega_{c}} \dfrac{\langle e |
    \hat{\phi}_{+} | e \rangle - \langle g | \hat{\phi}_{+} | g
    \rangle}{2} \approx - \dfrac{11 \sqrt{\omega_{0} \omega_{c}}
    \Delta^{2} E^{2}_{J} \lambda^{2} e^{- \lambda^{2}}}{3 E_{b} E_{c}}
    \sin f'. \label{eq:app:A.17d}
  \end{align}
  \label{eq:app:A.17}
\end{subequations}
\end{widetext}

\clearpage
\bibliography{reference}

\begin{thebibliography}{42}%
\makeatletter
\providecommand \@ifxundefined [1]{%
 \@ifx{#1\undefined}
}%
\providecommand \@ifnum [1]{%
 \ifnum #1\expandafter \@firstoftwo
 \else \expandafter \@secondoftwo
 \fi
}%
\providecommand \@ifx [1]{%
 \ifx #1\expandafter \@firstoftwo
 \else \expandafter \@secondoftwo
 \fi
}%
\providecommand \natexlab [1]{#1}%
\providecommand \enquote  [1]{``#1''}%
\providecommand \bibnamefont  [1]{#1}%
\providecommand \bibfnamefont [1]{#1}%
\providecommand \citenamefont [1]{#1}%
\providecommand \href@noop [0]{\@secondoftwo}%
\providecommand \href [0]{\begingroup \@sanitize@url \@href}%
\providecommand \@href[1]{\@@startlink{#1}\@@href}%
\providecommand \@@href[1]{\endgroup#1\@@endlink}%
\providecommand \@sanitize@url [0]{\catcode `\\12\catcode `\$12\catcode
  `\&12\catcode `\#12\catcode `\^12\catcode `\_12\catcode `\%12\relax}%
\providecommand \@@startlink[1]{}%
\providecommand \@@endlink[0]{}%
\providecommand \url  [0]{\begingroup\@sanitize@url \@url }%
\providecommand \@url [1]{\endgroup\@href {#1}{\urlprefix }}%
\providecommand \urlprefix  [0]{URL }%
\providecommand \Eprint [0]{\href }%
\providecommand \doibase [0]{http://dx.doi.org/}%
\providecommand \selectlanguage [0]{\@gobble}%
\providecommand \bibinfo  [0]{\@secondoftwo}%
\providecommand \bibfield  [0]{\@secondoftwo}%
\providecommand \translation [1]{[#1]}%
\providecommand \BibitemOpen [0]{}%
\providecommand \bibitemStop [0]{}%
\providecommand \bibitemNoStop [0]{.\EOS\space}%
\providecommand \EOS [0]{\spacefactor3000\relax}%
\providecommand \BibitemShut  [1]{\csname bibitem#1\endcsname}%
\let\auto@bib@innerbib\@empty
%</preamble>
\bibitem [{\citenamefont {Toyli}\ \emph {et~al.}(2016)\citenamefont {Toyli},
  \citenamefont {Eddins}, \citenamefont {Boutin}, \citenamefont {Puri},
  \citenamefont {Hover}, \citenamefont {Bolkhovsky}, \citenamefont {Oliver},
  \citenamefont {Blais},\ and\ \citenamefont {Siddiqi}}]{PhysRevX.6.031004}%
  \BibitemOpen
  \bibfield  {author} {\bibinfo {author} {\bibfnamefont {D.~M.}\ \bibnamefont
  {Toyli}}, \bibinfo {author} {\bibfnamefont {A.~W.}\ \bibnamefont {Eddins}},
  \bibinfo {author} {\bibfnamefont {S.}~\bibnamefont {Boutin}}, \bibinfo
  {author} {\bibfnamefont {S.}~\bibnamefont {Puri}}, \bibinfo {author}
  {\bibfnamefont {D.}~\bibnamefont {Hover}}, \bibinfo {author} {\bibfnamefont
  {V.}~\bibnamefont {Bolkhovsky}}, \bibinfo {author} {\bibfnamefont {W.~D.}\
  \bibnamefont {Oliver}}, \bibinfo {author} {\bibfnamefont {A.}~\bibnamefont
  {Blais}}, \ and\ \bibinfo {author} {\bibfnamefont {I.}~\bibnamefont
  {Siddiqi}},\ }\href {\doibase 10.1103/PhysRevX.6.031004} {\bibfield
  {journal} {\bibinfo  {journal} {Phys. Rev. X}\ }\textbf {\bibinfo {volume}
  {6}},\ \bibinfo {pages} {031004} (\bibinfo {year} {2016})}\BibitemShut
  {NoStop}%
\bibitem [{\citenamefont {Borja~Peropadre}\ and\ \citenamefont
  {Aspuru-Guzik}(2016)}]{BosonSampling2016}%
  \BibitemOpen
  \bibfield  {author} {\bibinfo {author} {\bibfnamefont {J.~H.}\ \bibnamefont
  {Borja~Peropadre}, \bibfnamefont {Gian Giacomo~Guerreschi}}\ and\ \bibinfo
  {author} {\bibfnamefont {A.}~\bibnamefont {Aspuru-Guzik}},\ }\href {\doibase
  10.1103/PhysRevLett.117.140505} {\bibfield  {journal} {\bibinfo  {journal}
  {Phys. Rev. Lett.}\ }\textbf {\bibinfo {volume} {117}},\ \bibinfo {pages}
  {140505} (\bibinfo {year} {2016})}\BibitemShut {NoStop}%
\bibitem [{\citenamefont {Zhou}\ \emph {et~al.}(2008)\citenamefont {Zhou},
  \citenamefont {Gong}, \citenamefont {Liu}, \citenamefont {Sun},\ and\
  \citenamefont {Nori}}]{PhysRevLett.101.100501}%
  \BibitemOpen
  \bibfield  {author} {\bibinfo {author} {\bibfnamefont {L.}~\bibnamefont
  {Zhou}}, \bibinfo {author} {\bibfnamefont {Z.~R.}\ \bibnamefont {Gong}},
  \bibinfo {author} {\bibfnamefont {Y.-x.}\ \bibnamefont {Liu}}, \bibinfo
  {author} {\bibfnamefont {C.~P.}\ \bibnamefont {Sun}}, \ and\ \bibinfo
  {author} {\bibfnamefont {F.}~\bibnamefont {Nori}},\ }\href {\doibase
  10.1103/PhysRevLett.101.100501} {\bibfield  {journal} {\bibinfo  {journal}
  {Phys. Rev. Lett.}\ }\textbf {\bibinfo {volume} {101}},\ \bibinfo {pages}
  {100501} (\bibinfo {year} {2008})}\BibitemShut {NoStop}%
\bibitem [{\citenamefont {He}\ \emph {et~al.}(2017)\citenamefont {He},
  \citenamefont {Zhu}, \citenamefont {Wang},\ and\ \citenamefont
  {Zhou}}]{He_singlephoton}%
  \BibitemOpen
  \bibfield  {author} {\bibinfo {author} {\bibfnamefont {Q.-K.}\ \bibnamefont
  {He}}, \bibinfo {author} {\bibfnamefont {W.}~\bibnamefont {Zhu}}, \bibinfo
  {author} {\bibfnamefont {Z.~H.}\ \bibnamefont {Wang}}, \ and\ \bibinfo
  {author} {\bibfnamefont {D.~L.}\ \bibnamefont {Zhou}},\ }\href
  {http://stacks.iop.org/0953-4075/50/i=14/a=145002} {\bibfield  {journal}
  {\bibinfo  {journal} {J. Phys. B: At. Mol. Opt. Phys.}\ }\textbf {\bibinfo
  {volume} {50}},\ \bibinfo {pages} {145002} (\bibinfo {year}
  {2017})}\BibitemShut {NoStop}%
\bibitem [{\citenamefont {Forn-D{\'\i}az}\ \emph {et~al.}(2017)\citenamefont
  {Forn-D{\'\i}az}, \citenamefont {Garc{\'\i}a-Ripoll}, \citenamefont
  {Peropadre}, \citenamefont {Orgiazzi}, \citenamefont {Yurtalan},
  \citenamefont {Belyansky}, \citenamefont {Wilson},\ and\ \citenamefont
  {Lupascu}}]{forn2017ultrastrong}%
  \BibitemOpen
  \bibfield  {author} {\bibinfo {author} {\bibfnamefont {P.}~\bibnamefont
  {Forn-D{\'\i}az}}, \bibinfo {author} {\bibfnamefont {J.}~\bibnamefont
  {Garc{\'\i}a-Ripoll}}, \bibinfo {author} {\bibfnamefont {B.}~\bibnamefont
  {Peropadre}}, \bibinfo {author} {\bibfnamefont {J.-L.}\ \bibnamefont
  {Orgiazzi}}, \bibinfo {author} {\bibfnamefont {M.}~\bibnamefont {Yurtalan}},
  \bibinfo {author} {\bibfnamefont {R.}~\bibnamefont {Belyansky}}, \bibinfo
  {author} {\bibfnamefont {C.}~\bibnamefont {Wilson}}, \ and\ \bibinfo {author}
  {\bibfnamefont {A.}~\bibnamefont {Lupascu}},\ }\href
  {https://www.nature.com/articles/nphys3905} {\bibfield  {journal} {\bibinfo
  {journal} {Nat. Phys.}\ }\textbf {\bibinfo {volume} {13}},\ \bibinfo {pages}
  {39} (\bibinfo {year} {2017})}\BibitemShut {NoStop}%
\bibitem [{\citenamefont {Forndiaz}\ \emph {et~al.}(2010)\citenamefont
  {Forndiaz}, \citenamefont {Lisenfeld}, \citenamefont {Marcos}, \citenamefont
  {Garciaripoll}, \citenamefont {Solano}, \citenamefont {Harmans},\ and\
  \citenamefont {Mooij}}]{forndiaz2010observation}%
  \BibitemOpen
  \bibfield  {author} {\bibinfo {author} {\bibfnamefont {P.}~\bibnamefont
  {Forndiaz}}, \bibinfo {author} {\bibfnamefont {J.}~\bibnamefont {Lisenfeld}},
  \bibinfo {author} {\bibfnamefont {D.}~\bibnamefont {Marcos}}, \bibinfo
  {author} {\bibfnamefont {J.~J.}\ \bibnamefont {Garciaripoll}}, \bibinfo
  {author} {\bibfnamefont {E.}~\bibnamefont {Solano}}, \bibinfo {author}
  {\bibfnamefont {C.~J. P.~M.}\ \bibnamefont {Harmans}}, \ and\ \bibinfo
  {author} {\bibfnamefont {J.~E.}\ \bibnamefont {Mooij}},\ }\href
  {https://journals.aps.org/prl/abstract/10.1103/PhysRevLett.105.237001}
  {\bibfield  {journal} {\bibinfo  {journal} {Phys. Rev. Lett.}\ }\textbf
  {\bibinfo {volume} {105}},\ \bibinfo {pages} {237001} (\bibinfo {year}
  {2010})}\BibitemShut {NoStop}%
\bibitem [{\citenamefont {Gu}\ \emph {et~al.}(2017)\citenamefont {Gu},
  \citenamefont {Kockum}, \citenamefont {Miranowicz}, \citenamefont {xi~Liu},\
  and\ \citenamefont {Nori}}]{GU20171}%
  \BibitemOpen
  \bibfield  {author} {\bibinfo {author} {\bibfnamefont {X.}~\bibnamefont
  {Gu}}, \bibinfo {author} {\bibfnamefont {A.~F.}\ \bibnamefont {Kockum}},
  \bibinfo {author} {\bibfnamefont {A.}~\bibnamefont {Miranowicz}}, \bibinfo
  {author} {\bibfnamefont {Y.}~\bibnamefont {xi~Liu}}, \ and\ \bibinfo {author}
  {\bibfnamefont {F.}~\bibnamefont {Nori}},\ }\href {\doibase
  https://doi.org/10.1016/j.physrep.2017.10.002} {\bibfield  {journal}
  {\bibinfo  {journal} {Physics Reports}\ }\textbf {\bibinfo {volume}
  {718-719}},\ \bibinfo {pages} {1 } (\bibinfo {year} {2017})}\BibitemShut
  {NoStop}%
\bibitem [{\citenamefont {Yoshihara}\ \emph {et~al.}(2017)\citenamefont
  {Yoshihara}, \citenamefont {Fuse}, \citenamefont {Ashhab}, \citenamefont
  {Kakuyanagi}, \citenamefont {Saito},\ and\ \citenamefont
  {Semba}}]{yoshihara2016superconducting}%
  \BibitemOpen
  \bibfield  {author} {\bibinfo {author} {\bibfnamefont {F.}~\bibnamefont
  {Yoshihara}}, \bibinfo {author} {\bibfnamefont {T.}~\bibnamefont {Fuse}},
  \bibinfo {author} {\bibfnamefont {S.}~\bibnamefont {Ashhab}}, \bibinfo
  {author} {\bibfnamefont {K.}~\bibnamefont {Kakuyanagi}}, \bibinfo {author}
  {\bibfnamefont {S.}~\bibnamefont {Saito}}, \ and\ \bibinfo {author}
  {\bibfnamefont {K.}~\bibnamefont {Semba}},\ }\href
  {https://www.nature.com/articles/nphys3906} {\bibfield  {journal} {\bibinfo
  {journal} {Nat. Phys.}\ }\textbf {\bibinfo {volume} {13}},\ \bibinfo {pages}
  {44} (\bibinfo {year} {2017})}\BibitemShut {NoStop}%
\bibitem [{\citenamefont {Peropadre}\ \emph
  {et~al.}(2013{\natexlab{a}})\citenamefont {Peropadre}, \citenamefont {Zueco},
  \citenamefont {Wulschner}, \citenamefont {Deppe}, \citenamefont {Marx},
  \citenamefont {Gross},\ and\ \citenamefont
  {Garciaripoll}}]{peropadre2013tunable}%
  \BibitemOpen
  \bibfield  {author} {\bibinfo {author} {\bibfnamefont {B.}~\bibnamefont
  {Peropadre}}, \bibinfo {author} {\bibfnamefont {D.}~\bibnamefont {Zueco}},
  \bibinfo {author} {\bibfnamefont {F.}~\bibnamefont {Wulschner}}, \bibinfo
  {author} {\bibfnamefont {F.}~\bibnamefont {Deppe}}, \bibinfo {author}
  {\bibfnamefont {A.}~\bibnamefont {Marx}}, \bibinfo {author} {\bibfnamefont
  {R.}~\bibnamefont {Gross}}, \ and\ \bibinfo {author} {\bibfnamefont {J.~J.}\
  \bibnamefont {Garciaripoll}},\ }\href
  {https://journals.aps.org/prb/abstract/10.1103/PhysRevB.87.134504} {\bibfield
   {journal} {\bibinfo  {journal} {Phys. Rev. B}\ }\textbf {\bibinfo {volume}
  {87}} (\bibinfo {year} {2013}{\natexlab{a}})}\BibitemShut {NoStop}%
\bibitem [{\citenamefont {Baust}\ \emph {et~al.}(2015)\citenamefont {Baust},
  \citenamefont {Hoffmann}, \citenamefont {Haeberlein}, \citenamefont
  {Schwarz}, \citenamefont {Eder}, \citenamefont {Menzel}, \citenamefont
  {Fedorov}, \citenamefont {Goetz}, \citenamefont {Wulschner}, \citenamefont
  {Xie} \emph {et~al.}}]{baust2015tunable}%
  \BibitemOpen
  \bibfield  {author} {\bibinfo {author} {\bibfnamefont {A.}~\bibnamefont
  {Baust}}, \bibinfo {author} {\bibfnamefont {E.}~\bibnamefont {Hoffmann}},
  \bibinfo {author} {\bibfnamefont {M.}~\bibnamefont {Haeberlein}}, \bibinfo
  {author} {\bibfnamefont {M.~J.}\ \bibnamefont {Schwarz}}, \bibinfo {author}
  {\bibfnamefont {P.}~\bibnamefont {Eder}}, \bibinfo {author} {\bibfnamefont
  {E.~P.}\ \bibnamefont {Menzel}}, \bibinfo {author} {\bibfnamefont {K.~G.}\
  \bibnamefont {Fedorov}}, \bibinfo {author} {\bibfnamefont {J.}~\bibnamefont
  {Goetz}}, \bibinfo {author} {\bibfnamefont {F.}~\bibnamefont {Wulschner}},
  \bibinfo {author} {\bibfnamefont {E.}~\bibnamefont {Xie}},  \emph {et~al.},\
  }\href {https://journals.aps.org/prb/abstract/10.1103/PhysRevB.91.014515}
  {\bibfield  {journal} {\bibinfo  {journal} {Phys. Rev. B}\ }\textbf {\bibinfo
  {volume} {91}} (\bibinfo {year} {2015})}\BibitemShut {NoStop}%
\bibitem [{\citenamefont {Wulschner}\ \emph {et~al.}(2016)\citenamefont
  {Wulschner}, \citenamefont {Goetz}, \citenamefont {Koessel}, \citenamefont
  {Hoffmann}, \citenamefont {Baust}, \citenamefont {Eder}, \citenamefont
  {Fischer}, \citenamefont {Haeberlein}, \citenamefont {Schwarz}, \citenamefont
  {Pernpeintner} \emph {et~al.}}]{wulschner2016tunable}%
  \BibitemOpen
  \bibfield  {author} {\bibinfo {author} {\bibfnamefont {F.}~\bibnamefont
  {Wulschner}}, \bibinfo {author} {\bibfnamefont {J.}~\bibnamefont {Goetz}},
  \bibinfo {author} {\bibfnamefont {F.~R.}\ \bibnamefont {Koessel}}, \bibinfo
  {author} {\bibfnamefont {E.}~\bibnamefont {Hoffmann}}, \bibinfo {author}
  {\bibfnamefont {A.}~\bibnamefont {Baust}}, \bibinfo {author} {\bibfnamefont
  {P.}~\bibnamefont {Eder}}, \bibinfo {author} {\bibfnamefont {M.}~\bibnamefont
  {Fischer}}, \bibinfo {author} {\bibfnamefont {M.}~\bibnamefont {Haeberlein}},
  \bibinfo {author} {\bibfnamefont {M.~J.}\ \bibnamefont {Schwarz}}, \bibinfo
  {author} {\bibfnamefont {M.}~\bibnamefont {Pernpeintner}},  \emph {et~al.},\
  }\href {https://link.springer.com/article/10.1140/epjqt/s40507-016-0048-2}
  {\bibfield  {journal} {\bibinfo  {journal} {EPJ Quantum Technology}\ }\textbf
  {\bibinfo {volume} {3}},\ \bibinfo {pages} {10} (\bibinfo {year}
  {2016})}\BibitemShut {NoStop}%
\bibitem [{\citenamefont {Kim}(2006)}]{PhysRevB.74.184501}%
  \BibitemOpen
  \bibfield  {author} {\bibinfo {author} {\bibfnamefont {M.~D.}\ \bibnamefont
  {Kim}},\ }\href {\doibase 10.1103/PhysRevB.74.184501} {\bibfield  {journal}
  {\bibinfo  {journal} {Phys. Rev. B}\ }\textbf {\bibinfo {volume} {74}},\
  \bibinfo {pages} {184501} (\bibinfo {year} {2006})}\BibitemShut {NoStop}%
\bibitem [{\citenamefont {Tong}\ \emph {et~al.}(2007)\citenamefont {Tong},
  \citenamefont {Singh}, \citenamefont {Kwek},\ and\ \citenamefont
  {Oh}}]{tong2007sufficiency}%
  \BibitemOpen
  \bibfield  {author} {\bibinfo {author} {\bibfnamefont {D.}~\bibnamefont
  {Tong}}, \bibinfo {author} {\bibfnamefont {K.}~\bibnamefont {Singh}},
  \bibinfo {author} {\bibfnamefont {L.~C.}\ \bibnamefont {Kwek}}, \ and\
  \bibinfo {author} {\bibfnamefont {C.~H.}\ \bibnamefont {Oh}},\ }\href
  {https://journals.aps.org/prl/abstract/10.1103/PhysRevLett.98.150402}
  {\bibfield  {journal} {\bibinfo  {journal} {Phys. Rev. Lett.}\ }\textbf
  {\bibinfo {volume} {98}},\ \bibinfo {pages} {150402} (\bibinfo {year}
  {2007})}\BibitemShut {NoStop}%
\bibitem [{\citenamefont {Martinis}\ \emph {et~al.}(2005)\citenamefont
  {Martinis}, \citenamefont {Cooper}, \citenamefont {McDermott}, \citenamefont
  {Steffen}, \citenamefont {Ansmann}, \citenamefont {Osborn}, \citenamefont
  {Cicak}, \citenamefont {Oh}, \citenamefont {Pappas}, \citenamefont
  {Simmonds},\ and\ \citenamefont {Yu}}]{PhysRevLett.95.210503}%
  \BibitemOpen
  \bibfield  {author} {\bibinfo {author} {\bibfnamefont {J.~M.}\ \bibnamefont
  {Martinis}}, \bibinfo {author} {\bibfnamefont {K.~B.}\ \bibnamefont
  {Cooper}}, \bibinfo {author} {\bibfnamefont {R.}~\bibnamefont {McDermott}},
  \bibinfo {author} {\bibfnamefont {M.}~\bibnamefont {Steffen}}, \bibinfo
  {author} {\bibfnamefont {M.}~\bibnamefont {Ansmann}}, \bibinfo {author}
  {\bibfnamefont {K.~D.}\ \bibnamefont {Osborn}}, \bibinfo {author}
  {\bibfnamefont {K.}~\bibnamefont {Cicak}}, \bibinfo {author} {\bibfnamefont
  {S.}~\bibnamefont {Oh}}, \bibinfo {author} {\bibfnamefont {D.~P.}\
  \bibnamefont {Pappas}}, \bibinfo {author} {\bibfnamefont {R.~W.}\
  \bibnamefont {Simmonds}}, \ and\ \bibinfo {author} {\bibfnamefont {C.~C.}\
  \bibnamefont {Yu}},\ }\href {\doibase 10.1103/PhysRevLett.95.210503}
  {\bibfield  {journal} {\bibinfo  {journal} {Phys. Rev. Lett.}\ }\textbf
  {\bibinfo {volume} {95}},\ \bibinfo {pages} {210503} (\bibinfo {year}
  {2005})}\BibitemShut {NoStop}%
\bibitem [{\citenamefont {Koch}\ \emph {et~al.}(2007)\citenamefont {Koch},
  \citenamefont {Terri}, \citenamefont {Gambetta}, \citenamefont {Houck},
  \citenamefont {Schuster}, \citenamefont {Majer}, \citenamefont {Blais},
  \citenamefont {Devoret}, \citenamefont {Girvin},\ and\ \citenamefont
  {Schoelkopf}}]{koch2007charge}%
  \BibitemOpen
  \bibfield  {author} {\bibinfo {author} {\bibfnamefont {J.}~\bibnamefont
  {Koch}}, \bibinfo {author} {\bibfnamefont {M.~Y.}\ \bibnamefont {Terri}},
  \bibinfo {author} {\bibfnamefont {J.}~\bibnamefont {Gambetta}}, \bibinfo
  {author} {\bibfnamefont {A.~A.}\ \bibnamefont {Houck}}, \bibinfo {author}
  {\bibfnamefont {D.}~\bibnamefont {Schuster}}, \bibinfo {author}
  {\bibfnamefont {J.}~\bibnamefont {Majer}}, \bibinfo {author} {\bibfnamefont
  {A.}~\bibnamefont {Blais}}, \bibinfo {author} {\bibfnamefont {M.~H.}\
  \bibnamefont {Devoret}}, \bibinfo {author} {\bibfnamefont {S.~M.}\
  \bibnamefont {Girvin}}, \ and\ \bibinfo {author} {\bibfnamefont {R.~J.}\
  \bibnamefont {Schoelkopf}},\ }\href
  {https://journals.aps.org/pra/abstract/10.1103/PhysRevA.76.042319} {\bibfield
   {journal} {\bibinfo  {journal} {Phys. Rev. A}\ }\textbf {\bibinfo {volume}
  {76}},\ \bibinfo {pages} {042319} (\bibinfo {year} {2007})}\BibitemShut
  {NoStop}%
\bibitem [{\citenamefont {Barends}\ \emph {et~al.}(2013)\citenamefont
  {Barends}, \citenamefont {Kelly}, \citenamefont {Megrant}, \citenamefont
  {Sank}, \citenamefont {Jeffrey}, \citenamefont {Chen}, \citenamefont {Yin},
  \citenamefont {Chiaro}, \citenamefont {Mutus}, \citenamefont {Neill} \emph
  {et~al.}}]{barends2013coherent}%
  \BibitemOpen
  \bibfield  {author} {\bibinfo {author} {\bibfnamefont {R.}~\bibnamefont
  {Barends}}, \bibinfo {author} {\bibfnamefont {J.}~\bibnamefont {Kelly}},
  \bibinfo {author} {\bibfnamefont {A.}~\bibnamefont {Megrant}}, \bibinfo
  {author} {\bibfnamefont {D.}~\bibnamefont {Sank}}, \bibinfo {author}
  {\bibfnamefont {E.}~\bibnamefont {Jeffrey}}, \bibinfo {author} {\bibfnamefont
  {Y.}~\bibnamefont {Chen}}, \bibinfo {author} {\bibfnamefont {Y.}~\bibnamefont
  {Yin}}, \bibinfo {author} {\bibfnamefont {B.}~\bibnamefont {Chiaro}},
  \bibinfo {author} {\bibfnamefont {J.}~\bibnamefont {Mutus}}, \bibinfo
  {author} {\bibfnamefont {C.}~\bibnamefont {Neill}},  \emph {et~al.},\ }\href
  {https://journals.aps.org/prl/abstract/10.1103/PhysRevLett.111.080502}
  {\bibfield  {journal} {\bibinfo  {journal} {Phys. Rev. Lett.}\ }\textbf
  {\bibinfo {volume} {111}},\ \bibinfo {pages} {080502} (\bibinfo {year}
  {2013})}\BibitemShut {NoStop}%
\bibitem [{\citenamefont {Inomata}\ \emph {et~al.}(2012)\citenamefont
  {Inomata}, \citenamefont {Yamamoto}, \citenamefont {Billangeon},
  \citenamefont {Nakamura},\ and\ \citenamefont {Tsai}}]{PhysRevB.86.140508}%
  \BibitemOpen
  \bibfield  {author} {\bibinfo {author} {\bibfnamefont {K.}~\bibnamefont
  {Inomata}}, \bibinfo {author} {\bibfnamefont {T.}~\bibnamefont {Yamamoto}},
  \bibinfo {author} {\bibfnamefont {P.-M.}\ \bibnamefont {Billangeon}},
  \bibinfo {author} {\bibfnamefont {Y.}~\bibnamefont {Nakamura}}, \ and\
  \bibinfo {author} {\bibfnamefont {J.~S.}\ \bibnamefont {Tsai}},\ }\href
  {\doibase 10.1103/PhysRevB.86.140508} {\bibfield  {journal} {\bibinfo
  {journal} {Phys. Rev. B}\ }\textbf {\bibinfo {volume} {86}},\ \bibinfo
  {pages} {140508} (\bibinfo {year} {2012})}\BibitemShut {NoStop}%
\bibitem [{\citenamefont {Mariantoni}\ \emph {et~al.}(2008)\citenamefont
  {Mariantoni}, \citenamefont {Deppe}, \citenamefont {Marx}, \citenamefont
  {Gross}, \citenamefont {Wilhelm},\ and\ \citenamefont
  {Solano}}]{PhysRevB.78.104508}%
  \BibitemOpen
  \bibfield  {author} {\bibinfo {author} {\bibfnamefont {M.}~\bibnamefont
  {Mariantoni}}, \bibinfo {author} {\bibfnamefont {F.}~\bibnamefont {Deppe}},
  \bibinfo {author} {\bibfnamefont {A.}~\bibnamefont {Marx}}, \bibinfo {author}
  {\bibfnamefont {R.}~\bibnamefont {Gross}}, \bibinfo {author} {\bibfnamefont
  {F.~K.}\ \bibnamefont {Wilhelm}}, \ and\ \bibinfo {author} {\bibfnamefont
  {E.}~\bibnamefont {Solano}},\ }\href {\doibase 10.1103/PhysRevB.78.104508}
  {\bibfield  {journal} {\bibinfo  {journal} {Phys. Rev. B}\ }\textbf {\bibinfo
  {volume} {78}},\ \bibinfo {pages} {104508} (\bibinfo {year}
  {2008})}\BibitemShut {NoStop}%
\bibitem [{\citenamefont {van~den Brink}\ \emph {et~al.}(2005)\citenamefont
  {van~den Brink}, \citenamefont {Berkley},\ and\ \citenamefont
  {Yalowsky}}]{1367-2630-7-1-230}%
  \BibitemOpen
  \bibfield  {author} {\bibinfo {author} {\bibfnamefont {A.~M.}\ \bibnamefont
  {van~den Brink}}, \bibinfo {author} {\bibfnamefont {A.~J.}\ \bibnamefont
  {Berkley}}, \ and\ \bibinfo {author} {\bibfnamefont {M.}~\bibnamefont
  {Yalowsky}},\ }\href {http://stacks.iop.org/1367-2630/7/i=1/a=230} {\bibfield
   {journal} {\bibinfo  {journal} {New J. Phys.}\ }\textbf {\bibinfo {volume}
  {7}},\ \bibinfo {pages} {230} (\bibinfo {year} {2005})}\BibitemShut {NoStop}%
\bibitem [{\citenamefont {Peropadre}\ \emph {et~al.}(2010)\citenamefont
  {Peropadre}, \citenamefont {Forn-D\'{\i}az}, \citenamefont {Solano},\ and\
  \citenamefont {Garc\'{\i}a-Ripoll}}]{PhysRevLett.105.023601}%
  \BibitemOpen
  \bibfield  {author} {\bibinfo {author} {\bibfnamefont {B.}~\bibnamefont
  {Peropadre}}, \bibinfo {author} {\bibfnamefont {P.}~\bibnamefont
  {Forn-D\'{\i}az}}, \bibinfo {author} {\bibfnamefont {E.}~\bibnamefont
  {Solano}}, \ and\ \bibinfo {author} {\bibfnamefont {J.~J.}\ \bibnamefont
  {Garc\'{\i}a-Ripoll}},\ }\href {\doibase 10.1103/PhysRevLett.105.023601}
  {\bibfield  {journal} {\bibinfo  {journal} {Phys. Rev. Lett.}\ }\textbf
  {\bibinfo {volume} {105}},\ \bibinfo {pages} {023601} (\bibinfo {year}
  {2010})}\BibitemShut {NoStop}%
\bibitem [{\citenamefont {Gambetta}\ \emph {et~al.}(2011)\citenamefont
  {Gambetta}, \citenamefont {Houck},\ and\ \citenamefont
  {Blais}}]{PhysRevLett.106.030502}%
  \BibitemOpen
  \bibfield  {author} {\bibinfo {author} {\bibfnamefont {J.~M.}\ \bibnamefont
  {Gambetta}}, \bibinfo {author} {\bibfnamefont {A.~A.}\ \bibnamefont {Houck}},
  \ and\ \bibinfo {author} {\bibfnamefont {A.}~\bibnamefont {Blais}},\ }\href
  {\doibase 10.1103/PhysRevLett.106.030502} {\bibfield  {journal} {\bibinfo
  {journal} {Phys. Rev. Lett.}\ }\textbf {\bibinfo {volume} {106}},\ \bibinfo
  {pages} {030502} (\bibinfo {year} {2011})}\BibitemShut {NoStop}%
\bibitem [{\citenamefont {Srinivasan}\ \emph {et~al.}(2011)\citenamefont
  {Srinivasan}, \citenamefont {Hoffman}, \citenamefont {Gambetta},\ and\
  \citenamefont {Houck}}]{srinivasan2011tunable}%
  \BibitemOpen
  \bibfield  {author} {\bibinfo {author} {\bibfnamefont {S.}~\bibnamefont
  {Srinivasan}}, \bibinfo {author} {\bibfnamefont {A.}~\bibnamefont {Hoffman}},
  \bibinfo {author} {\bibfnamefont {J.}~\bibnamefont {Gambetta}}, \ and\
  \bibinfo {author} {\bibfnamefont {A.}~\bibnamefont {Houck}},\ }\href
  {https://journals.aps.org/prl/abstract/10.1103/PhysRevLett.106.083601}
  {\bibfield  {journal} {\bibinfo  {journal} {Phys. Rev. Lett.}\ }\textbf
  {\bibinfo {volume} {106}},\ \bibinfo {pages} {083601} (\bibinfo {year}
  {2011})}\BibitemShut {NoStop}%
\bibitem [{\citenamefont {Bruschi}\ \emph {et~al.}(2013)\citenamefont
  {Bruschi}, \citenamefont {Lee},\ and\ \citenamefont
  {Fuentes}}]{bruschi2013time}%
  \BibitemOpen
  \bibfield  {author} {\bibinfo {author} {\bibfnamefont {D.~E.}\ \bibnamefont
  {Bruschi}}, \bibinfo {author} {\bibfnamefont {A.~R.}\ \bibnamefont {Lee}}, \
  and\ \bibinfo {author} {\bibfnamefont {I.}~\bibnamefont {Fuentes}},\ }\href
  {http://stacks.iop.org/1751-8121/46/i=16/a=165303} {\bibfield  {journal}
  {\bibinfo  {journal} {J. Phys. A: Math. Theor.}\ }\textbf {\bibinfo {volume}
  {46}},\ \bibinfo {pages} {165303} (\bibinfo {year} {2013})}\BibitemShut
  {NoStop}%
\bibitem [{\citenamefont {Mezzacapo}\ \emph {et~al.}(2014)\citenamefont
  {Mezzacapo}, \citenamefont {Lamata}, \citenamefont {Filipp},\ and\
  \citenamefont {Solano}}]{mezzacapo2014many}%
  \BibitemOpen
  \bibfield  {author} {\bibinfo {author} {\bibfnamefont {A.}~\bibnamefont
  {Mezzacapo}}, \bibinfo {author} {\bibfnamefont {L.}~\bibnamefont {Lamata}},
  \bibinfo {author} {\bibfnamefont {S.}~\bibnamefont {Filipp}}, \ and\ \bibinfo
  {author} {\bibfnamefont {E.}~\bibnamefont {Solano}},\ }\href {\doibase
  10.1103/PhysRevLett.113.050501} {\bibfield  {journal} {\bibinfo  {journal}
  {Phys. Rev. Lett.}\ }\textbf {\bibinfo {volume} {113}},\ \bibinfo {pages}
  {050501} (\bibinfo {year} {2014})}\BibitemShut {NoStop}%
\bibitem [{\citenamefont {Lu}\ \emph {et~al.}(2017)\citenamefont {Lu},
  \citenamefont {Chakram}, \citenamefont {Leung}, \citenamefont {Earnest},
  \citenamefont {Naik}, \citenamefont {Huang}, \citenamefont {Groszkowski},
  \citenamefont {Kapit}, \citenamefont {Koch},\ and\ \citenamefont
  {Schuster}}]{PhysRevLett.119.150502}%
  \BibitemOpen
  \bibfield  {author} {\bibinfo {author} {\bibfnamefont {Y.}~\bibnamefont
  {Lu}}, \bibinfo {author} {\bibfnamefont {S.}~\bibnamefont {Chakram}},
  \bibinfo {author} {\bibfnamefont {N.}~\bibnamefont {Leung}}, \bibinfo
  {author} {\bibfnamefont {N.}~\bibnamefont {Earnest}}, \bibinfo {author}
  {\bibfnamefont {R.~K.}\ \bibnamefont {Naik}}, \bibinfo {author}
  {\bibfnamefont {Z.}~\bibnamefont {Huang}}, \bibinfo {author} {\bibfnamefont
  {P.}~\bibnamefont {Groszkowski}}, \bibinfo {author} {\bibfnamefont
  {E.}~\bibnamefont {Kapit}}, \bibinfo {author} {\bibfnamefont
  {J.}~\bibnamefont {Koch}}, \ and\ \bibinfo {author} {\bibfnamefont {D.~I.}\
  \bibnamefont {Schuster}},\ }\href {\doibase 10.1103/PhysRevLett.119.150502}
  {\bibfield  {journal} {\bibinfo  {journal} {Phys. Rev. Lett.}\ }\textbf
  {\bibinfo {volume} {119}},\ \bibinfo {pages} {150502} (\bibinfo {year}
  {2017})}\BibitemShut {NoStop}%
\bibitem [{\citenamefont {Kyaw}\ \emph {et~al.}(2015)\citenamefont {Kyaw},
  \citenamefont {Felicetti}, \citenamefont {Romero}, \citenamefont {Solano},\
  and\ \citenamefont {Kwek}}]{kyaw2015scalable}%
  \BibitemOpen
  \bibfield  {author} {\bibinfo {author} {\bibfnamefont {T.~H.}\ \bibnamefont
  {Kyaw}}, \bibinfo {author} {\bibfnamefont {S.}~\bibnamefont {Felicetti}},
  \bibinfo {author} {\bibfnamefont {G.}~\bibnamefont {Romero}}, \bibinfo
  {author} {\bibfnamefont {E.}~\bibnamefont {Solano}}, \ and\ \bibinfo {author}
  {\bibfnamefont {L.~C.}\ \bibnamefont {Kwek}},\ }\href {\doibase
  10.1038/srep08621} {\bibfield  {journal} {\bibinfo  {journal} {Scientific
  Reports}\ }\textbf {\bibinfo {volume} {5}},\ \bibinfo {pages} {8621}
  (\bibinfo {year} {2015})}\BibitemShut {NoStop}%
\bibitem [{\citenamefont {Felicetti}\ \emph {et~al.}(2015)\citenamefont
  {Felicetti}, \citenamefont {Sab\'{\i}n}, \citenamefont {Fuentes},
  \citenamefont {Lamata}, \citenamefont {Romero},\ and\ \citenamefont
  {Solano}}]{PhysRevB.92.064501}%
  \BibitemOpen
  \bibfield  {author} {\bibinfo {author} {\bibfnamefont {S.}~\bibnamefont
  {Felicetti}}, \bibinfo {author} {\bibfnamefont {C.}~\bibnamefont
  {Sab\'{\i}n}}, \bibinfo {author} {\bibfnamefont {I.}~\bibnamefont {Fuentes}},
  \bibinfo {author} {\bibfnamefont {L.}~\bibnamefont {Lamata}}, \bibinfo
  {author} {\bibfnamefont {G.}~\bibnamefont {Romero}}, \ and\ \bibinfo {author}
  {\bibfnamefont {E.}~\bibnamefont {Solano}},\ }\href {\doibase
  10.1103/PhysRevB.92.064501} {\bibfield  {journal} {\bibinfo  {journal} {Phys.
  Rev. B}\ }\textbf {\bibinfo {volume} {92}},\ \bibinfo {pages} {064501}
  (\bibinfo {year} {2015})}\BibitemShut {NoStop}%
\bibitem [{\citenamefont {Garciaalvarez}\ \emph {et~al.}(2017)\citenamefont
  {Garciaalvarez}, \citenamefont {Felicetti}, \citenamefont {Rico},
  \citenamefont {Solano},\ and\ \citenamefont
  {Sabin}}]{garciaalvarez2017entanglement}%
  \BibitemOpen
  \bibfield  {author} {\bibinfo {author} {\bibfnamefont {L.}~\bibnamefont
  {Garciaalvarez}}, \bibinfo {author} {\bibfnamefont {S.}~\bibnamefont
  {Felicetti}}, \bibinfo {author} {\bibfnamefont {E.}~\bibnamefont {Rico}},
  \bibinfo {author} {\bibfnamefont {E.}~\bibnamefont {Solano}}, \ and\ \bibinfo
  {author} {\bibfnamefont {C.}~\bibnamefont {Sabin}},\ }\href {\doibase
  10.1038/s41598-017-00770-z} {\bibfield  {journal} {\bibinfo  {journal}
  {Scientific Reports}\ }\textbf {\bibinfo {volume} {7}},\ \bibinfo {pages}
  {657} (\bibinfo {year} {2017})}\BibitemShut {NoStop}%
\bibitem [{\citenamefont {Devoret}(1995)}]{DevoretQuanFluc}%
  \BibitemOpen
  \bibfield  {author} {\bibinfo {author} {\bibfnamefont {M.}~\bibnamefont
  {Devoret}},\ }\href
  {http://www.copilot.caltech.edu/documents/260-les_houches_devoret_quantum_fluctuations_electrical_circuits_1997.pdf}
  {\bibfield  {journal} {\bibinfo  {journal} {Les Houches, Session LXIII}\
  }\textbf {\bibinfo {volume} {7}},\ \bibinfo {pages} {351} (\bibinfo {year}
  {1995})}\BibitemShut {NoStop}%
\bibitem [{\citenamefont {Peropadre}\ \emph
  {et~al.}(2013{\natexlab{b}})\citenamefont {Peropadre}, \citenamefont {Zueco},
  \citenamefont {Porras},\ and\ \citenamefont
  {Garc\'{\i}a-Ripoll}}]{PhysRevLett.111.243602}%
  \BibitemOpen
  \bibfield  {author} {\bibinfo {author} {\bibfnamefont {B.}~\bibnamefont
  {Peropadre}}, \bibinfo {author} {\bibfnamefont {D.}~\bibnamefont {Zueco}},
  \bibinfo {author} {\bibfnamefont {D.}~\bibnamefont {Porras}}, \ and\ \bibinfo
  {author} {\bibfnamefont {J.~J.}\ \bibnamefont {Garc\'{\i}a-Ripoll}},\ }\href
  {\doibase 10.1103/PhysRevLett.111.243602} {\bibfield  {journal} {\bibinfo
  {journal} {Phys. Rev. Lett.}\ }\textbf {\bibinfo {volume} {111}},\ \bibinfo
  {pages} {243602} (\bibinfo {year} {2013}{\natexlab{b}})}\BibitemShut
  {NoStop}%
\bibitem [{\citenamefont {Sakurai}\ and\ \citenamefont
  {Napolitano}(2011)}]{Sakurai2011Modern}%
  \BibitemOpen
  \bibfield  {author} {\bibinfo {author} {\bibfnamefont {J.~J.}\ \bibnamefont
  {Sakurai}}\ and\ \bibinfo {author} {\bibfnamefont {J.~J.}\ \bibnamefont
  {Napolitano}},\ }\href
  {https://www.pearsonhighered.com/program/Sakurai-Modern-Quantum-Mechanics-2nd-Edition/PGM160720.html}
  {\emph {\bibinfo {title} {{Modern Quantum Mechanics, 2nd Edition}}}}\
  (\bibinfo  {publisher} {Addison-Wesley \& Pearson},\ \bibinfo {year} {2011})\
  pp.\ \bibinfo {pages} {285--304}\BibitemShut {NoStop}%
\bibitem [{\citenamefont {Braak}(2011)}]{PhysRevLett.107.100401}%
  \BibitemOpen
  \bibfield  {author} {\bibinfo {author} {\bibfnamefont {D.}~\bibnamefont
  {Braak}},\ }\href {\doibase 10.1103/PhysRevLett.107.100401} {\bibfield
  {journal} {\bibinfo  {journal} {Phys. Rev. Lett.}\ }\textbf {\bibinfo
  {volume} {107}},\ \bibinfo {pages} {100401} (\bibinfo {year}
  {2011})}\BibitemShut {NoStop}%
\bibitem [{\citenamefont {Ithier}\ \emph {et~al.}(2005)\citenamefont {Ithier},
  \citenamefont {Collin}, \citenamefont {Joyez}, \citenamefont {Meeson},
  \citenamefont {Vion}, \citenamefont {Esteve}, \citenamefont {Chiarello},
  \citenamefont {Shnirman}, \citenamefont {Makhlin}, \citenamefont {Schriefl}
  \emph {et~al.}}]{ithier2005decoherence}%
  \BibitemOpen
  \bibfield  {author} {\bibinfo {author} {\bibfnamefont {G.}~\bibnamefont
  {Ithier}}, \bibinfo {author} {\bibfnamefont {E.}~\bibnamefont {Collin}},
  \bibinfo {author} {\bibfnamefont {P.}~\bibnamefont {Joyez}}, \bibinfo
  {author} {\bibfnamefont {P.}~\bibnamefont {Meeson}}, \bibinfo {author}
  {\bibfnamefont {D.}~\bibnamefont {Vion}}, \bibinfo {author} {\bibfnamefont
  {D.}~\bibnamefont {Esteve}}, \bibinfo {author} {\bibfnamefont
  {F.}~\bibnamefont {Chiarello}}, \bibinfo {author} {\bibfnamefont
  {A.}~\bibnamefont {Shnirman}}, \bibinfo {author} {\bibfnamefont
  {Y.}~\bibnamefont {Makhlin}}, \bibinfo {author} {\bibfnamefont
  {J.}~\bibnamefont {Schriefl}},  \emph {et~al.},\ }\href
  {https://journals.aps.org/prb/abstract/10.1103/PhysRevB.72.134519} {\bibfield
   {journal} {\bibinfo  {journal} {Phys. Rev. B}\ }\textbf {\bibinfo {volume}
  {72}},\ \bibinfo {pages} {134519} (\bibinfo {year} {2005})}\BibitemShut
  {NoStop}%
\bibitem [{\citenamefont {Schoelkopf}\ \emph {et~al.}(2003)\citenamefont
  {Schoelkopf}, \citenamefont {Clerk}, \citenamefont {Girvin}, \citenamefont
  {Lehnert},\ and\ \citenamefont {Devoret}}]{schoelkopf2003noise}%
  \BibitemOpen
  \bibfield  {author} {\bibinfo {author} {\bibfnamefont {R.~J.}\ \bibnamefont
  {Schoelkopf}}, \bibinfo {author} {\bibfnamefont {A.}~\bibnamefont {Clerk}},
  \bibinfo {author} {\bibfnamefont {S.}~\bibnamefont {Girvin}}, \bibinfo
  {author} {\bibfnamefont {K.~W.}\ \bibnamefont {Lehnert}}, \ and\ \bibinfo
  {author} {\bibfnamefont {M.}~\bibnamefont {Devoret}},\ }in\ \href {\doibase
  10.1117/12.488922} {\emph {\bibinfo {booktitle} {Proc. SPIE}}},\ Vol.\
  \bibinfo {volume} {5115}\ (\bibinfo {organization} {International Society for
  Optics and Photonics},\ \bibinfo {year} {2003})\ pp.\ \bibinfo {pages}
  {356--377}\BibitemShut {NoStop}%
\bibitem [{\citenamefont {Bouchiat}\ \emph {et~al.}(1998)\citenamefont
  {Bouchiat}, \citenamefont {Vion}, \citenamefont {Joyez}, \citenamefont
  {Esteve},\ and\ \citenamefont {Devoret}}]{bouchiat1998quantum}%
  \BibitemOpen
  \bibfield  {author} {\bibinfo {author} {\bibfnamefont {V.}~\bibnamefont
  {Bouchiat}}, \bibinfo {author} {\bibfnamefont {D.}~\bibnamefont {Vion}},
  \bibinfo {author} {\bibfnamefont {P.}~\bibnamefont {Joyez}}, \bibinfo
  {author} {\bibfnamefont {D.}~\bibnamefont {Esteve}}, \ and\ \bibinfo {author}
  {\bibfnamefont {M.}~\bibnamefont {Devoret}},\ }\href
  {http://iopscience.iop.org/article/10.1238/Physica.Topical.076a00165/meta}
  {\bibfield  {journal} {\bibinfo  {journal} {Physica Scripta}\ }\textbf
  {\bibinfo {volume} {1998}},\ \bibinfo {pages} {165} (\bibinfo {year}
  {1998})}\BibitemShut {NoStop}%
\bibitem [{\citenamefont {Nakamura}\ \emph {et~al.}(1999)\citenamefont
  {Nakamura}, \citenamefont {Pashkin},\ and\ \citenamefont
  {Tsai}}]{nakamura1999coherent}%
  \BibitemOpen
  \bibfield  {author} {\bibinfo {author} {\bibfnamefont {Y.}~\bibnamefont
  {Nakamura}}, \bibinfo {author} {\bibfnamefont {Y.~A.}\ \bibnamefont
  {Pashkin}}, \ and\ \bibinfo {author} {\bibfnamefont {J.~S.}\ \bibnamefont
  {Tsai}},\ }\href {\doibase 10.1038/19718} {\bibfield  {journal} {\bibinfo
  {journal} {Nature}\ }\textbf {\bibinfo {volume} {398}},\ \bibinfo {pages}
  {786} (\bibinfo {year} {1999})}\BibitemShut {NoStop}%
\bibitem [{\citenamefont {Friedman}\ \emph {et~al.}(2000)\citenamefont
  {Friedman}, \citenamefont {Patel}, \citenamefont {Chen}, \citenamefont
  {Tolpygo},\ and\ \citenamefont {Lukens}}]{friedman2000jr}%
  \BibitemOpen
  \bibfield  {author} {\bibinfo {author} {\bibfnamefont {J.~R.}\ \bibnamefont
  {Friedman}}, \bibinfo {author} {\bibfnamefont {V.}~\bibnamefont {Patel}},
  \bibinfo {author} {\bibfnamefont {W.}~\bibnamefont {Chen}}, \bibinfo {author}
  {\bibfnamefont {S.~K.}\ \bibnamefont {Tolpygo}}, \ and\ \bibinfo {author}
  {\bibfnamefont {J.~E.}\ \bibnamefont {Lukens}},\ }\href
  {https://www.nature.com/articles/35017505} {\bibfield  {journal} {\bibinfo
  {journal} {Nature (London)}\ }\textbf {\bibinfo {volume} {406}},\ \bibinfo
  {pages} {43} (\bibinfo {year} {2000})}\BibitemShut {NoStop}%
\bibitem [{\citenamefont {Der~Wal}\ \emph {et~al.}(2000)\citenamefont
  {Der~Wal}, \citenamefont {Haar}, \citenamefont {Wilhelm}, \citenamefont
  {Schouten}, \citenamefont {Harmans}, \citenamefont {Orlando}, \citenamefont
  {Lloyd},\ and\ \citenamefont {Mooij}}]{derwal2000quantum}%
  \BibitemOpen
  \bibfield  {author} {\bibinfo {author} {\bibfnamefont {C.~H.~V.}\
  \bibnamefont {Der~Wal}}, \bibinfo {author} {\bibfnamefont {A.~C. J.~T.}\
  \bibnamefont {Haar}}, \bibinfo {author} {\bibfnamefont {F.~K.}\ \bibnamefont
  {Wilhelm}}, \bibinfo {author} {\bibfnamefont {R.~N.}\ \bibnamefont
  {Schouten}}, \bibinfo {author} {\bibfnamefont {C.~J. P.~M.}\ \bibnamefont
  {Harmans}}, \bibinfo {author} {\bibfnamefont {T.~P.}\ \bibnamefont
  {Orlando}}, \bibinfo {author} {\bibfnamefont {S.}~\bibnamefont {Lloyd}}, \
  and\ \bibinfo {author} {\bibfnamefont {J.~E.}\ \bibnamefont {Mooij}},\ }\href
  {http://science.sciencemag.org/content/290/5492/773} {\bibfield  {journal}
  {\bibinfo  {journal} {Science}\ }\textbf {\bibinfo {volume} {290}},\ \bibinfo
  {pages} {773} (\bibinfo {year} {2000})}\BibitemShut {NoStop}%
\bibitem [{\citenamefont {Paladino}\ \emph {et~al.}(2014)\citenamefont
  {Paladino}, \citenamefont {Galperin}, \citenamefont {Falci},\ and\
  \citenamefont {Altshuler}}]{paladino20141}%
  \BibitemOpen
  \bibfield  {author} {\bibinfo {author} {\bibfnamefont {E.}~\bibnamefont
  {Paladino}}, \bibinfo {author} {\bibfnamefont {Y.}~\bibnamefont {Galperin}},
  \bibinfo {author} {\bibfnamefont {G.}~\bibnamefont {Falci}}, \ and\ \bibinfo
  {author} {\bibfnamefont {B.}~\bibnamefont {Altshuler}},\ }\href
  {https://journals.aps.org/rmp/abstract/10.1103/RevModPhys.86.361} {\bibfield
  {journal} {\bibinfo  {journal} {Rev. Mod. Phys.}\ }\textbf {\bibinfo {volume}
  {86}},\ \bibinfo {pages} {361} (\bibinfo {year} {2014})}\BibitemShut
  {NoStop}%
\bibitem [{\citenamefont {Yoshihara}\ \emph {et~al.}(2006)\citenamefont
  {Yoshihara}, \citenamefont {Harrabi}, \citenamefont {Niskanen}, \citenamefont
  {Nakamura},\ and\ \citenamefont {Tsai}}]{yoshihara2006decoherence}%
  \BibitemOpen
  \bibfield  {author} {\bibinfo {author} {\bibfnamefont {F.}~\bibnamefont
  {Yoshihara}}, \bibinfo {author} {\bibfnamefont {K.}~\bibnamefont {Harrabi}},
  \bibinfo {author} {\bibfnamefont {A.}~\bibnamefont {Niskanen}}, \bibinfo
  {author} {\bibfnamefont {Y.}~\bibnamefont {Nakamura}}, \ and\ \bibinfo
  {author} {\bibfnamefont {J.}~\bibnamefont {Tsai}},\ }\href
  {https://journals.aps.org/prl/abstract/10.1103/PhysRevLett.97.167001}
  {\bibfield  {journal} {\bibinfo  {journal} {Phys. Rev. Lett.}\ }\textbf
  {\bibinfo {volume} {97}},\ \bibinfo {pages} {167001} (\bibinfo {year}
  {2006})}\BibitemShut {NoStop}%
\bibitem [{\citenamefont {Van~Harlingen}\ \emph {et~al.}(2004)\citenamefont
  {Van~Harlingen}, \citenamefont {Robertson}, \citenamefont {Plourde},
  \citenamefont {Reichardt}, \citenamefont {Crane},\ and\ \citenamefont
  {Clarke}}]{van2004decoherence}%
  \BibitemOpen
  \bibfield  {author} {\bibinfo {author} {\bibfnamefont {D.}~\bibnamefont
  {Van~Harlingen}}, \bibinfo {author} {\bibfnamefont {T.}~\bibnamefont
  {Robertson}}, \bibinfo {author} {\bibfnamefont {B.}~\bibnamefont {Plourde}},
  \bibinfo {author} {\bibfnamefont {P.}~\bibnamefont {Reichardt}}, \bibinfo
  {author} {\bibfnamefont {T.}~\bibnamefont {Crane}}, \ and\ \bibinfo {author}
  {\bibfnamefont {J.}~\bibnamefont {Clarke}},\ }\href
  {https://journals.aps.org/prb/abstract/10.1103/PhysRevB.70.064517} {\bibfield
   {journal} {\bibinfo  {journal} {Phys. Rev. B}\ }\textbf {\bibinfo {volume}
  {70}},\ \bibinfo {pages} {064517} (\bibinfo {year} {2004})}\BibitemShut
  {NoStop}%
\bibitem [{\citenamefont {Zorin}\ \emph {et~al.}(1996)\citenamefont {Zorin},
  \citenamefont {Ahlers}, \citenamefont {Niemeyer}, \citenamefont {Weimann},
  \citenamefont {Wolf}, \citenamefont {Krupenin},\ and\ \citenamefont
  {Lotkhov}}]{PhysRevB.53.13682}%
  \BibitemOpen
  \bibfield  {author} {\bibinfo {author} {\bibfnamefont {A.~B.}\ \bibnamefont
  {Zorin}}, \bibinfo {author} {\bibfnamefont {F.-J.}\ \bibnamefont {Ahlers}},
  \bibinfo {author} {\bibfnamefont {J.}~\bibnamefont {Niemeyer}}, \bibinfo
  {author} {\bibfnamefont {T.}~\bibnamefont {Weimann}}, \bibinfo {author}
  {\bibfnamefont {H.}~\bibnamefont {Wolf}}, \bibinfo {author} {\bibfnamefont
  {V.~A.}\ \bibnamefont {Krupenin}}, \ and\ \bibinfo {author} {\bibfnamefont
  {S.~V.}\ \bibnamefont {Lotkhov}},\ }\href {\doibase
  10.1103/PhysRevB.53.13682} {\bibfield  {journal} {\bibinfo  {journal} {Phys.
  Rev. B}\ }\textbf {\bibinfo {volume} {53}},\ \bibinfo {pages} {13682}
  (\bibinfo {year} {1996})}\BibitemShut {NoStop}%
\end{thebibliography}%

\end{document}